\documentclass[twocolumn]{aastex62}

\newcommand{\teff} {\ensuremath{T_{\rm eff}}}
\newcommand{\mjup} {\ensuremath{M_{\rm Jup}}}
\newcommand{\Msun}{\ensuremath{M_{\odot}}}

\usepackage{comment}
\usepackage[figure,figure*]{hypcap}

\graphicspath{{./}{figures/}}
\received{January 1, 2018}
\revised{January 7, 2018}
\accepted{\today}
\submitjournal{ApJ}

\shorttitle{Substellar objects in the Orion Nebula}
\shortauthors{Robberto et al.}

\begin{document}

\title{HST survey of the Orion Nebula Cluster in the H$_2$O 1.4~$\mu$m absorption band:\\
I. A census of substellar and planetary mass objects
}

\correspondingauthor{Massimo Robberto}
\email{robberto@stsci.edu}

\author[0000-0002-0786-7307]{Massimo Robberto}
\affiliation{Space Telescope Science Institute, 3700 San Martin Dr, Baltimore, MD 21218, USA}
\altaffiliation{
\begin{minipage}{28em}
Based on observations made with the NASA/ESA {\it Hubble
Space Telescope}, obtained at the Space Telescope Science Institute, which
is operated by the Association of Universities for Research in Astronomy, 
Inc., under NASA contract NAS 5-26555.  These observations are associated
with program GO-13826.
\end{minipage}
}
\affiliation{Johns Hopkins University, 3400 N. Charles Street, Baltimore, MD 21218, USA}

\author[0000-0002-5581-2896]{Mario Gennaro}
\affiliation{Space Telescope Science Institute, 3700 San Martin Dr, Baltimore, MD 21218, USA}

\author[0000-0002-5980-4287]{Maria Giulia Ubeira Gabellini}
\affiliation{Dipartimento di Fisica, Universit{\`a} Degli Studi di Milano, Via Giovanni Celoria, 16, 20133 Milano, MI, Italy}
\affiliation{European Southern Observatory, Karl-Schwarzschild-Str. 2, 85784 Garching, Germany}

\author{Lynne A. Hillenbrand}
\affil{California Institute of Technology, MC 249-17, 1200 East California Blvd, Pasadena CA 91125, USA}

\author[0000-0003-4196-0617]{Camilla Pacifici}
\affiliation{Space Telescope Science Institute, 3700 San Martin Dr, Baltimore, MD 21218, USA}


\author{Leonardo Ubeda}
\affiliation{Space Telescope Science Institute, 3700 San Martin Dr, Baltimore, MD 21218, USA}


\author[0000-0002-5306-4089]{Morten Andersen}
\affil{Gemini Observatory, NSF’s National Optical-Infrared Astronomy Research Laboratory, Casilla 603, La Serena, Chile}

\author[0000-0002-7129-3002]{Travis Barman}
\affil{Lunar and Planetary Laboratory,
1629 E University Blvd,
Tucson AZ 85721-0092, USA }

\author[0000-0003-3858-637X]{Andrea Bellini}
\affiliation{Space Telescope Science Institute, 3700 San Martin Dr, Baltimore, MD 21218, USA}


\author{Nicola Da Rio}
\affil{University of Virginia, Department of Astronomy,
P.O. Box 400325, 530 McCormick Road,
Charlottesville, VA 22904-4325, USA}

\author[0000-0001-9336-2825]{Selma E. de Mink}
\affiliation{Center for Astrophysics, Harvard-Smithsonian, 60 Garden Street, Cambridge, MA 02138, USA}
\affiliation{Anton Pannekoek Institute for Astronomy, University of Amsterdam, Science Park 904, 1098XH Amsterdam, The Netherlands}

\author[0000-0002-2357-7692]{Giuseppe Lodato}
\affiliation{Dipartimento di Fisica, Universit{\`a} Degli Studi di Milano, Via Giovanni Celoria, 16, 20133 Milano, MI, Italy}

\author[0000-0003-3562-262X]{Carlo Felice Manara}
\affiliation{European Southern Observatory,
Karl-Schwarzschild-Str. 2,
85748 Garching, Germany}


\author[0000-0003-2599-2459]{Imants Platais}
\affiliation{Johns Hopkins University, 3400 N. Charles Street, Baltimore, MD 21218, USA}

\author[0000-0003-3818-408X]{Laurent Pueyo}
\affiliation{Space Telescope Science Institute, 3700 San Martin Dr, Baltimore, MD 21218, USA}





\author[0000-0002-1652-420X]{Giovanni Maria Strampelli}
\affiliation{Johns Hopkins University, 3400 N. Charles Street, Baltimore, MD 21218, USA}
\affiliation{Space Telescope Science Institute, 3700 San Martin Dr, Baltimore, MD 21218, USA}
\affiliation{Department of Astrophysics, University of La Laguna, Av. Astrofísico Francisco Sánchez, 38200 San Cristóbal de La Laguna, Tenerife, Canary Islands, Spain}

\author[0000-0002-3389-9142]{Jonathan C. Tan}
\affil{Dept. of Space, Earth \& Environment, Chalmers University of Technology, Gothenburg, Sweden}
\affiliation{Dept. of Astronomy, University of Virginia, Charlottesville, Virginia, USA}

\author[0000-0003-1859-3070]{L Testi} 
\affiliation{European Southern Observatory,
Karl-Schwarzschild-Str. 2,
85748 Garching, Germany}

\begin{abstract}
In order to obtain a complete census of the stellar and sub-stellar population, down to a few \mjup~ in the $\sim1$~Myr old Orion Nebula Cluster, we used the infrared channel of the Wide Field Camera 3 of the Hubble Space Telescope with the F139M and F130N  filters. These bandpasses correspond to  the  $1.4~\mu$m H$_2$O  absorption feature  and an adjacent line-free continuum region.  Out of  4,504 detected sources, 3,352 (about 75\%) appear fainter than $m_{130}=14$ (Vega mag) in the F130N filter, a brightness corresponding to the hydrogen-burning limit mass ($M\simeq 0.072~\Msun$) at $\sim 1$~Myr. Of these, however, only 742 sources have a negative F130M-139N color index, indicative of the presence of H$_2$O vapor in absorption, and can  therefore be classified as  bona-fide M and L dwarfs, with effective temperatures  $T\lesssim 2850$~K at an assumed  1~Myr cluster age. On our color-magnitude diagram, this population of sources with H$_2$O absorption appears clearly distinct from the larger background population of highly reddened stars and galaxies with positive F130M-F139N color index, and can be traced down to the sensitivity limit of our survey, $m_{130}\simeq 21.5$, corresponding to a 1~Myr old $\simeq 3$ \mjup\, planetary mass object
under about 2 magnitudes of visual extinction. Theoretical models of the BT-Settl family predicting substellar isochrones of 1, 2 and 3~Myr {down to $\sim 1~\mjup$} fail to reproduce the observed H$_2$O color index at $M\lesssim 20$~\mjup. We perform a Bayesian analysis to determine extinction, mass and effective temperature of each sub-stellar member of our sample, together with its membership probability. 
\end{abstract}


\section{Introduction}


With $\sim$2,000 members, the Orion Nebula Cluster (ONC) is the richest young ($\simeq1-2$~Myr) cluster within 2~kpc from the Sun \citep{2003ARA&A..41...57L, doi:10.1146/annurev-astro-081309-130834}. The cluster spans the full range of spectral types, from massive Main-Sequence OB-type stars to substellar brown dwarfs. Due to its modest distance from the Sun \cite[$d\approx400$~pc][]{2018AJ....156...84K, refId0} and to the relatively low foreground extinction \citep[$A_V\sim1$][]{2011AA...533A..38S}, the ONC can be studied 
in great detail. 
By characterizing the main parameters of the multitude of ONC pre-main-sequence stars, several studies have determined the main cluster properties, such as the structure, dynamics, star formation rate and initial mass function.
The reviews of \cite{2008hsf1.book..483M} and \cite{2008hsf1.book..544O} summarize the status of our knowledge before 2008. Further progress has been made in the last decade through photometric \citep[e.g.,][]{2010ApJ...722.1092D, 2012ApJ...748...14D, 2011A&A...534A..10A, 2013ApJS..207...10R, 2016MNRAS.461.1734D} and spectroscopic surveys \citep[e.g.,][]{2013AJ....146...85H, 2016ApJ...818...59D, 2014ApJ...782....8I}. 

Even if the ONC is usually regarded as the standard benchmark for comparative studies of young clusters, our knowledge of this system is not complete.
Obtaining a complete census of the full population down to the minimum mass for opacity-limited fragmentation \citep[$M\sim1-10~\mjup\rm$, see][and reference therein]{2018haex.bookE..92B} has not yet been achieved.
One challenge is that the ONC is still partially embedded within its parental molecular cloud, carved by the ionizing radiation of the most massive stars, $\theta^1$Ori-C in particular. The molecular cloud provides a backdrop that mitigates background confusion, but its column density, and thus opacity, is not homogeneous \citep[e.g.][]{2011AA...533A..38S,2018ApJS..236...25K}: background objects can contaminate source counts, especially at infrared wavelengths where the extinction is lower. Moreover, the side of the molecular cloud facing us is illuminated by the brightest ONC stars; the high surface brightness limits the sensitivity and completeness especially in the central regions where the cluster stellar density reaches its peak. The non-uniformity of the foreground extinction  \citep{2019ApJ...881..130A}, the contamination from foreground pre-main-sequence (PMS) stars associated with older Orion associations \citep{2014A&A...564A..29B}, the presence of circumstellar emission 
\citep[proplyds, jets,e.g.][]{1996AJ....111..846O,2000AJ....119.2919B,2008AJ....136.2136R}, 
 photospheric emission {\citep[accretion, e.g.][]{2004ApJ...606..952R,2012ApJ...755..154M}}, {local extinction by envelopes and circumstellar disks with processed grains \citep{2016ApJ...826...16E}}, source variability {\citep{2002A&A...396..513H,2012ApJ...753..149M}}, etc. hamper in many cases the capability of assigning a well defined spectral type. Conventional imaging surveys carried out in broad-band filters are effective at finding low-mass candidates, but firm separation from contaminants is problematic \citep{2017AJ....154..256C}.
Spectroscopic confirmation is generally needed to disentangle low mass members against spurious contaminants.

Spectroscopic surveys of the ONC have identified about 60 brown dwarfs (BDs) \citep{2004ApJ...610.1045S, 2007MNRAS.381.1077R, 2009MNRAS.392..817W}, a relatively low number. Photometric studies \citep{2000ApJ...540..236H, 2012ApJ...748...14D} confirm that the population of low-mass sources is rather small. Other groups, however, find evidence of a rich BD population \citep{2002ApJ...573..366M, 2005MNRAS.361..211L, 2016MNRAS.461.1734D}. 

To overcome the difficulty of determining the precise shape of the low-mass Initial Mass Function (IMF), coarser diagnostics have been introduced. An often-used parameterization is the simple star/brown dwarf ratio $R$ between the number of objects in the two mass bins from $0.08$ to $1.0~\Msun$ and from 0.03 to $0.08 \Msun$. This has been adopted by \cite{2011A&A...534A..10A}, who analysed HST/NICMOS data  taken in the $JH$ bands over a small  ($\simeq20\%$), non-contiguous fraction of the
ONC field down to an unprecedented depth $H\sim 22$ \citep{2013ApJS..207...10R}. They derived ratios ranging from $R = 3.7$ at the center down to $R\sim 1$ in the periphery, i.e. a significant depletion of substellar objects at the cluster core. At these faint sensitivity levels, however, contamination becomes a major issue as broad-band $JH$ photometry cannot directly discriminate between cluster members and the multitude of background sources. In the case of the ONC, the accuracy of the correction applied to account for background contamination is limited by the non-uniformity of the nebular background, both in terms of brightness and extinction.

To avoid these limitations, \cite{2010ApJ...722.1092D, 2012ApJ...748...14D} adopted a safer approach based on a direct measure of (\teff). M dwarfs have been originally characterized by the onset of the molecular TiO absorption feature at visible wavelengths \citep{1943assw.book.....M}. By using CCD photometry  in the TiO molecular bands at $\sim7700$ \AA, present in the atmosphere of PMS stars later than $\sim$M2 and brown dwarfs earlier than L2, \citep{Kirkpatrick_1999} these authors placed 1750 sources in the HR diagram with minimal contamination from background sources, deriving a cluster IMF with a lower fraction of substellar objects than that found by \cite{2011A&A...534A..10A}. Observationally, the drawback in this case is that
these observations were performed in the visible and from the ground. Therefore they are not sensitive to the faintest and/or more embedded sources, in particular the low mass tail of brown dwarfs and planetary-mass objects probed more efficiently, but with higher confusion, by \cite{2011A&A...534A..10A}. 

\begin{figure*}[htb!]
\begin{center}
\includegraphics[width=\textwidth,angle=0]{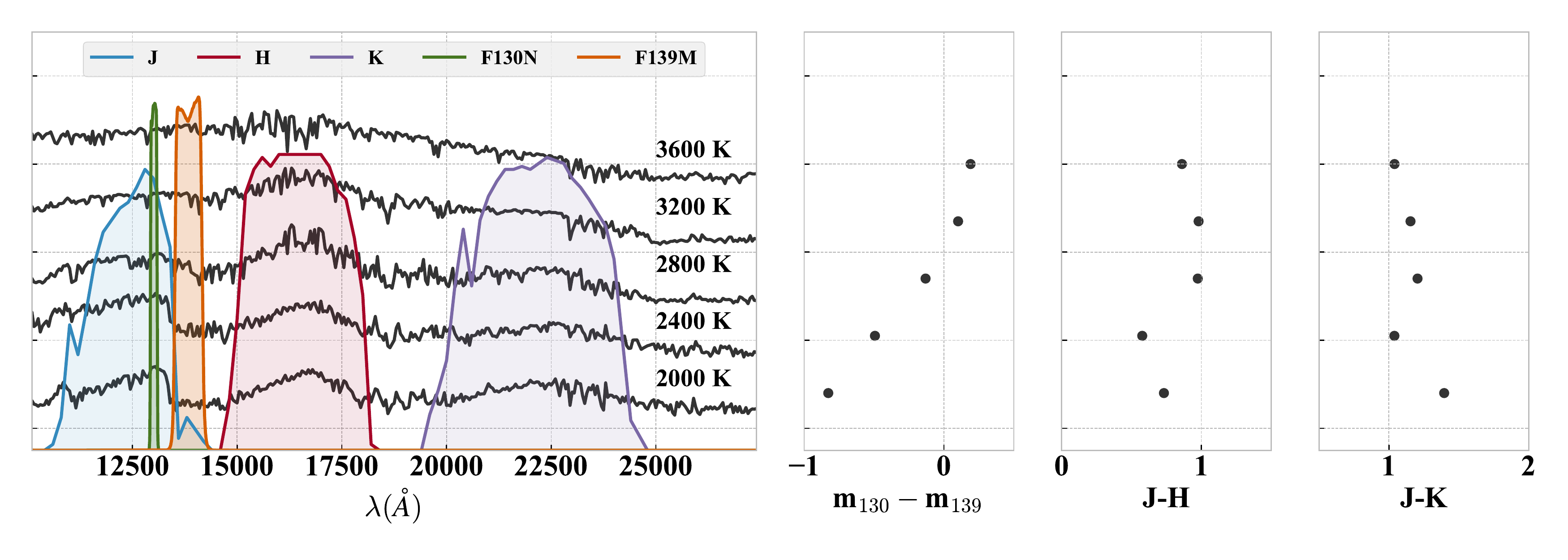}
\caption{Left: Theoretical near-IR spectra (BT-Settl) for five different values of $T_{eff}$, with the transmission profiles of the broad-band $J$, $H$, $K$ bandpasses and our narrower F130N and F139M filters of WFC3; Right: color indexes derived from our H$_2$O-sensitive filter combination of F130N-F139M, and from the the $J-H$ and $H-K$ magnitude difference, for the same $T_{eff}$ values assumed in left plot (top to bottom: 3600~K, 3200~K, 2800~K, 2400~K and 2000~K, {roughly corresponding to $M=0.7, 0.2, 0.06, 0.015$ and $0.006$\Msun, see Appendix A)}. Surface gravities are $\log(g)\simeq3.5$. The diagnostic power of the H$_2$O index is apparent from the systematic and large growth of the index towards cooler temperatures. \label{Fig:H2Oband} } 
\end{center}
\end{figure*}

The results of these two studies suggest that to firmly identify the substellar members of the ONC it would be ideal to combine the strengths of the two methods: the depth of HST IR observations and the diagnostic power of an absorption feature sensitive to \teff\, (and therefore to the
mass) of very-low-mass objects. 
\cite{2000ApJ...541..977N} were able to determine the spectral type and reddening for every M-type star in the field of young cluster IC~348 using a photometric index based on the strength of the 1.9~$\mu$m water band. This feature was accessible at that time with the HST/NICMOS instrument, but is beyond the spectral range of the more advanced HST/WFC3 that has superseded NICMOS.
A similar feature that remains uniquely accessible by HST/WFC3 is the $1.345~\mu$m H$_2$O absorption band that on the ground separates the $J$ and $H$ bands 
. 
The depth of the $1.345~\mu$m band is known to correlate with \teff\, in brown dwarfs  \citep[][Figure~\ref{Fig:H2Oband}]{1994MNRAS.267..413J, 1999A&AS..135...41D}, and remains prominent down to very low temperatures, \teff$\lesssim500\,$K, when the outer planetary atmospheres are cool enough to condense water vapor into clouds \citep{2001ApJ...556..872A, 2003ApJ...596..587B, 2014ApJ...787...78M}.
{According to the models, the F139M (covering the H$_2$O feature) and the F130N (adjacent continuum) filters of the Wide Field Camera 3 should provide a nearly reddening-free photometric index sensitive to \teff\, (thus, a model-dependent mass) that at 6~\mjup\, can be traced up to $A_V\simeq40$~mag. This because the H$_2$O index changes by just $\sim 0.023$~mag/$A_V$, the two infrared bandpasses being close in wavelength}. Due to its sensitivity to very cool objects, the H$_2$O index effectively discriminates young cluster members from reddened but intrinsically warmer background objects \citep{2000ApJ...535..325W, 2003ApJ...593.1074G}. 
%
 

In  Section  2 we present the observations, detailing the survey strategy, data processing and analysis and the final photometric catalog. In Section 3 we illustrate our completeness analysis. {In Section 4 we present the main results, i.e. the capability of our measures to discriminate effectively between low mass objects of the ONC and background stars, the relation between our water index and the effective temperature, the derived source parameters and an analysis of the effect of contamination on the luminosity and mass functions. We present our strategy to overcome the limitation of the current models and derive a semi-empirical isochrone in the two Appendixes.} A full derivation of the ONC low-mass IMF, based on an extensive Bayesian analysis our our data, is presented in the second paper of this series by Gennaro \& Robberto (2020, ApJ in press, hereafter Paper~II).

\section{Observations}
\begin{figure*}[t]
\begin{center}
\includegraphics[width=\textwidth, trim={0 1.5cm 0 1cm}]{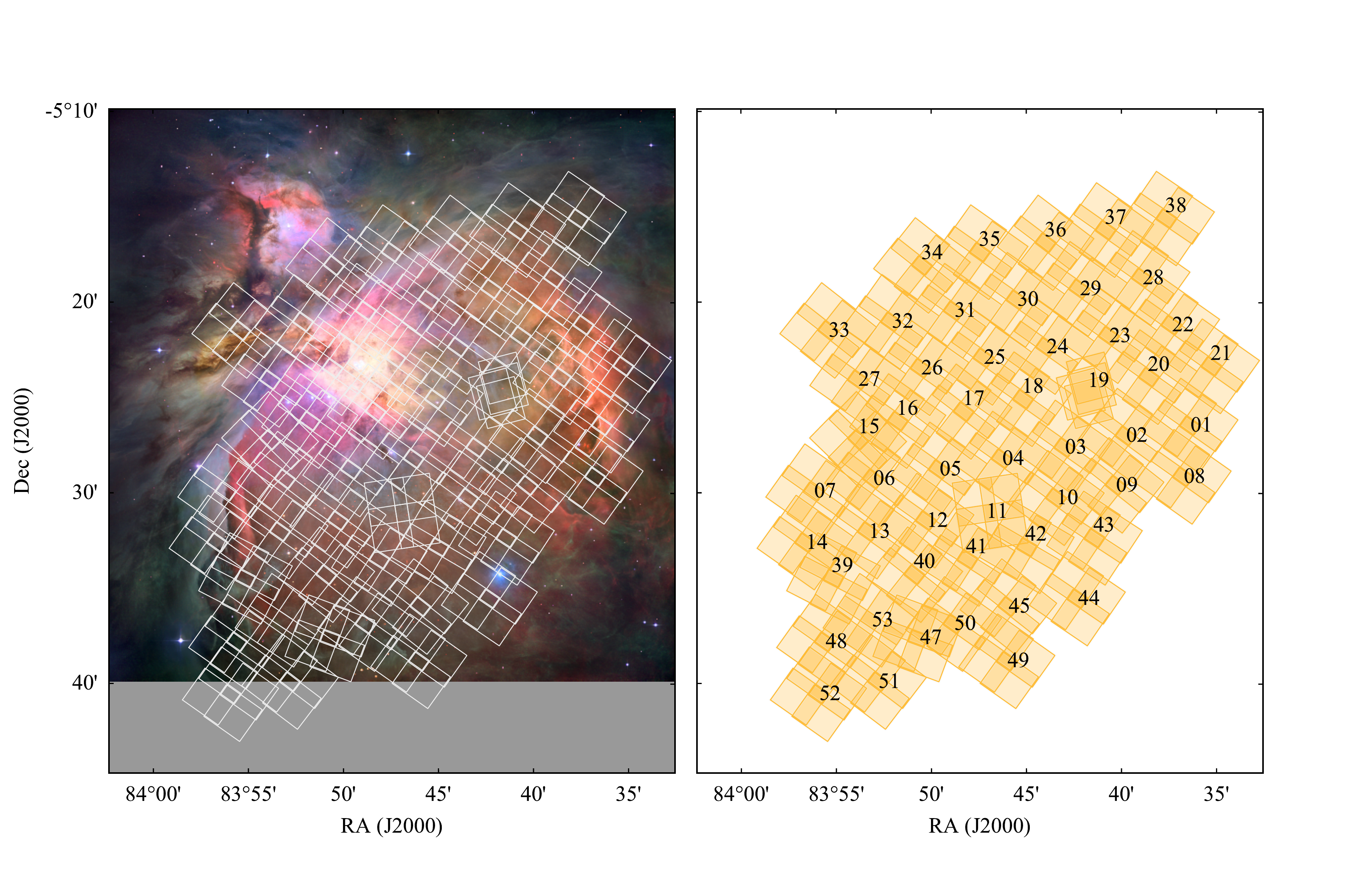}
\caption{Field coverage of the HST GO-13826 observations presented in this paper. The large $2\times$ dither pattern of each visit is apparent, together with the irregular  configurations of the visits lacking suitable guide stars for nominal orientations.   \label{Fig:FieldCoverage} } 
\end{center}
\end{figure*}

The data presented in this paper were obtained as part of the Cycle 22 HST Treasury Program {\sl The Orion Nebula Cluster as a Paradigm of Star Formation''} (GO-13826, P.I. M. Robberto). Observations were carried out between 8 February and 29 November 2015 using the ACS-WFC and WFC3-IR cameras in coordinated parallel mode. For the WFC3-IR survey we used the F139M  and F130N filters. The F139M filter  \citep[labeled as {H$_2$O/CH$_4$} filter in the WFC3 Instrument Handbook][]{Dressel2014} is a medium-band filter with central wavelength $\lambda_c = 1.383~\mu$m 
and  width $\Delta\lambda=64.3$~nm.
The F130N, nominally indicated as {Paschen-continuum} filter, is a narrow-band filter with $\lambda_c = 1.300~\mu$m  and $\Delta\lambda=15.6$~nm.  We adopted this  filter, instead of the  standard  {H$_2$O/CH$_4$-continuum}  $F127M$ filter of WFC3, in order to measure the photospheric continuum without contamination from the $\lambda=1.28~\mu$m  Paschen-$\beta$ line,  prominent  in the Orion Nebula region.

\subsection{Survey Strategy}
Our observing strategy was driven by the goal of imaging with both cameras the same area covered by the 2005 ONC Treasury Program \citep[HST GO-10246][]{2013ApJS..207...10R}  with adequate depth in the near-IR to identify all sources previously detected in the ACS F850LP filter. To facilitate the analysis of proper motion (Platais et al., 2020, ApJ submitted), our original plan was to start each visit with the same pointing and orientation used by the first survey, replicating the exposures obtained 10 years before with the same instrument and filter. In fact, since the tiling pattern used in 2005 had 50\% overlap between visits, replicating only half of the original sequence of 104 visits was enough to cover the full field. In practice, various updates implemented to increase the accuracy of the HST guide star catalog have reduced the number of candidate guide stars in the Orion Nebula field.
Therefore, for a number of pointings we could not find suitable guide stars. In order to cover those areas, we had to change the orientation of the telescope. This is the reason why 21 visits have roll angle different from the desired 100 or 280 degrees and appear more randomly distributed on the otherwise regular mosaic pattern (see Figure~\ref{Fig:FieldCoverage}).

Within each visit, corresponding to an HST orbit with about 45 min of target visibility, we maximized the WFC3-IR field coverage by taking four exposures per filter with large offsets; specifically, we pointed the telescope to the four vertices of a square with 120 arcsec diagonal.
This is the maximum distance allowed by the HST for a pointing maneuver that keeps a single guide star in use. The pointing overhead in this case is only 60 seconds; moving further away forces the acquisition of a new guide star, that would require 6 minutes of overheads and  strongly degrade the survey efficiency. The four pointings at the vertex of the inscribed square produce a square mosaic covering an area $\sim 2.8$ times larger than the nominal WFC3-IR field of view, i.e., 13 arcmin$^{2}$ instead of the original 4.64 arcmin$^{2}$. About $35\%$ of the field, in the central overlapping cross, is covered by two or more frames.
Depending on the pointing, adjacent visits also have some degree of overlap, which increases the fraction of field covered at least twice.  

With four dithered images in two filters per visit, our 52 visits WFC3-IR survey turns out to be composed by 416 individual images, covering overall a field of 0.135 deg$^{2}$.


The  sequence of  8 exposures per  visit, with parameters listed  in  Table~\ref{tbl-1},  alternates  between the  two filters to reduce the number of wheel moves.  To  pack 8  ramps within  a single orbit, we  used SAMP-SEQ=SPARS100 with NSAMP=4, corresponding to about 304~s of total integration time;  however, two of the  four  exposures in the F130N filter had to be executed with SAMP-SEQ=SPARS50 and 5 frames, corresponding to about 204~s of integration time, in order to remain within the total duration of the visibility window.

\begin{deluxetable}{cccccc}
\tabletypesize{\scriptsize}
\tablecaption{Observing Sequence per Visit\label{tbl-1}}
\tablewidth{0pt}
\tablehead{
\colhead{Filter} & 
 \colhead{SAMP-SEQ} & 
  \colhead{NSAMP} & 
   \colhead{Exp. Time (s)} & 
    \colhead{$\Delta X\arcsec$} & 
     \colhead{$\Delta Y\arcsec$} 
}
\startdata
F130N & SPARS100 & 4 & 303.933 & 0    &   0    \\
F139M & SPARS100 & 4 & 303.933 & 0    &   0    \\
F139M & SPARS100 & 4 & 303.933 & 0    &   120    \\
F130N & SPARS100 & 4 & 303.933 & 0    &   120    \\
F130N & SPARS50 & 5 & 203.934 & 60    &   60    \\
F139M & SPARS100 & 4 & 303.933 & 60    &   60    \\
F139M & SPARS100 & 4 & 303.933 & -60    &   60    \\
F130N & SPARS50 & 5 & 203.934 & -60    &   60    \\
\enddata
\tablecomments{In the order, from left to right: column 1) HST/WFC3 filter denomination; 2) WFC timing sequence; 3) Number of non-destructive samples per integration; 4) Total exposure time per integration; 5 and 6) Typical commanded offset per dither move along the detector $X$ and $Y$ axes}
\end{deluxetable}

\subsection{Data Processing}
Each image was initially processed using the standard WFC3 pipeline, that delivers count rate images corrected for flat-field and calibrated in units of electrons/second (suffix {\rm FLT}). 
To derive the position of each source and reconstruct the full mosaic, the 416 images had to be distortion-corrected, co-aligned and placed at the absolute sky position, relying on the telescope offset information for the dithered exposures and on the ACS source catalog of \cite{2013ApJS..207...10R} for the large area mosaic.

We did not rely on the accuracy of the astrometric information in the native FITS headers released by the HST archive, as they were limited by the uncertainties on the absolute coordinates of the Guide Star Catalog and the accuracy and stability of the distance between the FGS guiders and the WFC3 focal plane. 
The astrometric information stored in the FITS header, therefore, was modified to recover the true absolute pointing of the telescope. 
We started with the images taken in the F139M filter, having higher signal-to-noise, combining first the four images of each visit into single tiles. For this operation we relied on the relative offsets commanded to the telescope, executed with precision of a few milli-arcsec, since the four images had been taken using the same guide star. We thus produced 52 distortion corrected tiles, each one a mosaic of $2\times2$ dithered frames. 

The next step consisted  in assigning to each tile the  correct absolute astrometry, updating the  WCS  parameters stored in the  FITS headers. 
To this purpose we used the ACS source catalogue of \cite{2013ApJS..207...10R}, as it provides a rich  ($n=2674$) and dense dataset of sources detected in both the F775W ACS filter and our F139M filter. Each tile was thus registered with a relative accuracy of about $1/10^{th}$ of a ACS pixel, or 5~mas. The absolute precision is limited by the accuracy of the ACS source catalog. A preliminary comparison with the GAIA coordinates shows discrepancies of the order of 50~mas or 1 ACS pixel, accurate enough for cross-matching against any other existing photometric catalog. An extensive astrometric analysis of the ACS dataset based on Gaia DR2 coordiantes will be presented in Platais et al. 2020.

\begin{figure*}[htb!]
\begin{center}
\includegraphics[width=180 mm]{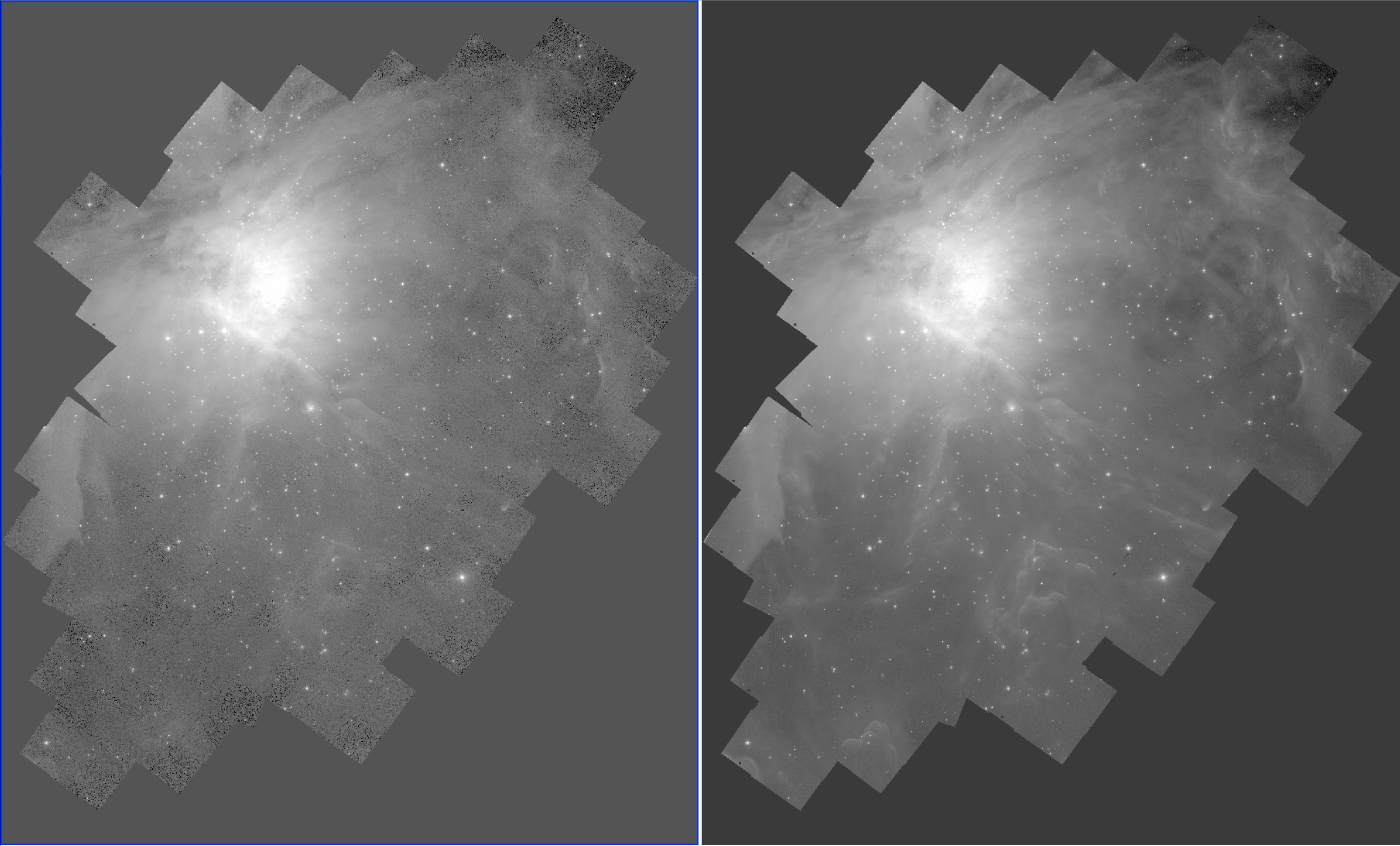}
\caption{HST/WFC3-IR mosaics produced by our survey in the two filters F130N (left) and F139M (right). The two images, taken in adjacent bandpasses, are nearly indistinguishable except for the higher noise in the narrower F130N bandbass, as shown e.g. in the lower part of the field. The Trapezium region is the saturated area in both images.} \label{2bigfilt}
\end{center}
\end{figure*}

Having registered our mosaic tiles on an absolute coordinate system, we used  AstroDrizzle to create a full mosaic in the F139M filter first,  
then in  the F130N mosaic by adopting the same astrometric information. The two mosaics in Figure~\ref{2bigfilt} are therefore registered to each other with accuracy of a few milliarseconds, or about 1/100$^{th}$ of a WFC3/IR pixel.

\subsection{Photometry\label{sec:photometry}}
We derived the photometry of each source using a combination of aperture and PSF photometry. 
For PSF photometry we used DOLPHOT \citep{2000PASP..112.1383D}, that uses a model PSF based on the HST TinyTim simulator \citep{2011SPIE.8127E..0JK}
Since the HST PSF is not perfectly stable, DOLPHOT refines its shape in an iterative way, through comparison with the empirical one extracted from isolated stars in the field. This can be problematic in the case of the ONC, due to the presence of circumstellar structures, both in emission or in absorption such as photoionized proplyds, dark disks, jets, as well as the scarcity of suitable sources in selected fields. For this reason, we paid special attention to  parameters like the $\chi^2$ returned by DOLPHOT to reject problematic sources. We found that in many cases the $\chi^2$ values were worse for the F130N filter than for the F139M PSFs; this is probably due to the fact that the F139M PSF are avaialble 
with a correction (``Anderson core'') that is currently not available for the narrow-band filter\footnote{See \href{url}{http://americano.dolphinsim.com/dolphot/} for the latest version of DOLPHOT code and tables.}
\citep[see e.g.][for a discussion of the limitations of the TinyTim  point-spread function models used for
PSF photometry]{2012ApJS..200...18D}. Realizing that we could rely also on the more accurate but less precise aperture photometry, we adopted the  general strategy of averaging multi-aperture (3 and 15 pixels) and PSF photometry when the  values were coincident within the errors and the reduced $\chi^2$ of the PSF photometry was close to 1; otherwise, we used only aperture photometry. { For the final photometric catalog, we averaged the values for the sources detected multiple times.}
Saturation starts to affect photometry at $m_{139}\simeq9.5$ and $m_{130}\simeq10.9$ (Vega magnitudes). 
To reliably identify only real stars (point source)s, we vetted the source catalog automatically created by DOLPHOT.
After rejecting the sources having DOLPHOT parameters (sharpness, roundness, $\chi^2$) and magnitude errors deviating from the median values of their magnitude bins by several sigmas,
we performed a visual inspection of each source to uncover and reject residual artifacts such as the diffraction spikes produced by the brighter stars that DOLPHOT occasionally resolves in  individual sources. We also rejected binaries found by DOLPHOT with separation $<0.26$~arcsec (2 pixels),  using in these cases only the values reported for the brightest source. A study of the binary population is the subject of Paper-III of this series (Strampelli et al., Ap.J. in press). 

A last type of artifact we had to deal with is image persistence, which is the latent images of bright sources caused by the delayed release of trapped electrons in IR detectors \citep{2008SPIE.7021E..0JS}. 
Assuming that persistence was affecting only the data taken within one orbit, meaning that at our flux levels the decay time is generally shorter than the duration of the earth occultation, we flagged the pixels falling under the core of a bright star in each of the following exposures, so that the first pair (both filters) of exposures in an orbit was immune while  the last pair was the most contaminated. Visual inspection at the expected positions showed that persistence artifacts were clearly present in our images, spread over the full field due to the large dithers, but they had been almost totally rejected from our source catalog due to the abnormal parameters of the PSF fitting, in both filters.

\begin{figure}[htb!]
\begin{center}
\includegraphics[width=90mm,angle=0]{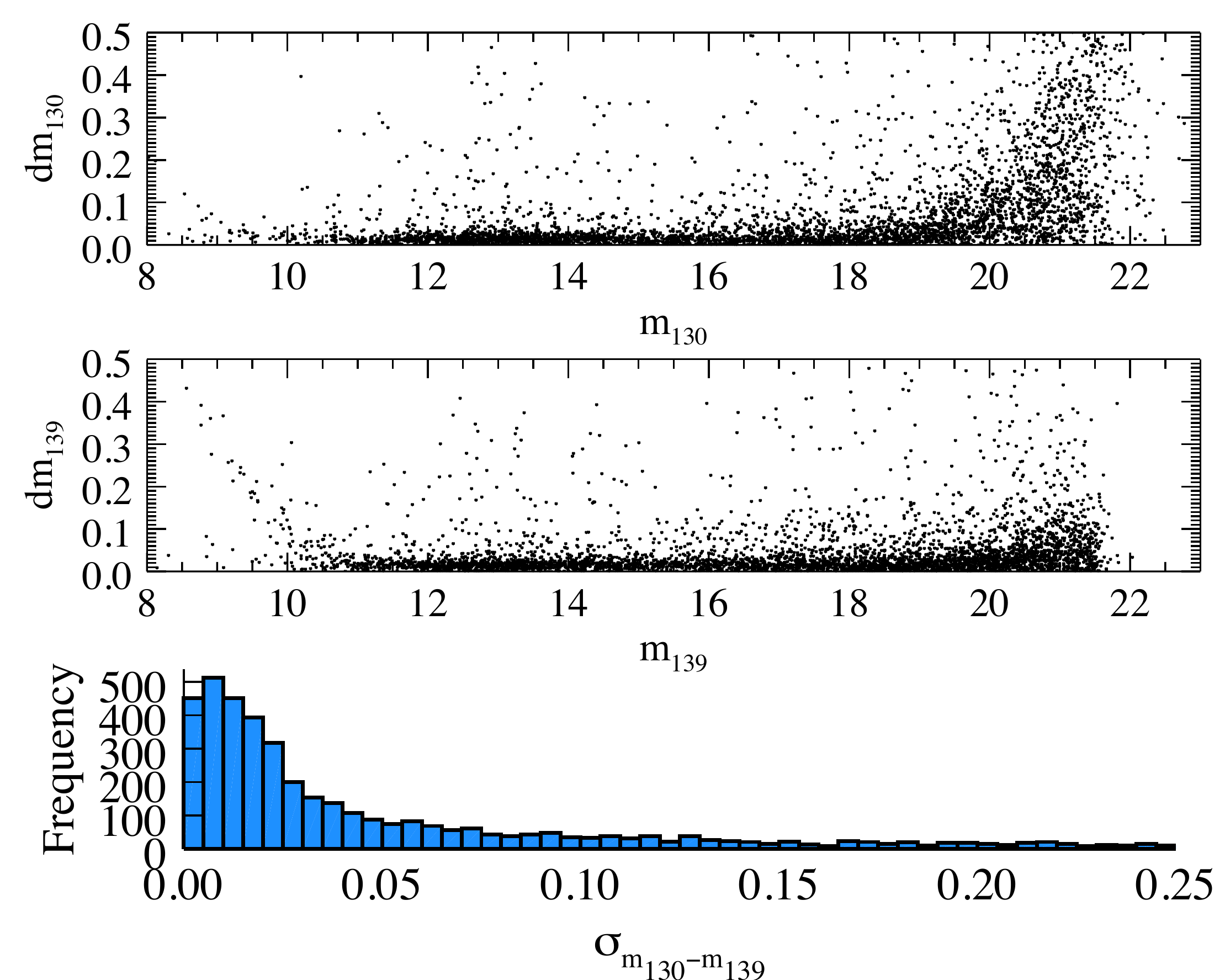}
\caption{Magnitude error as a function of the $m_{130}$ (top) and $m_{139}$ (middle) magnitudes, and histogram of the uncertainties in the $(m_{130}-m_{139})$ color index (bottom). \label{Fig:3plots}}
\end{center}
\end{figure}

\begin{figure*}[htb!]
\begin{center}
\includegraphics[width=150mm,angle=0,trim={0 2.5cm 0 5cm}]{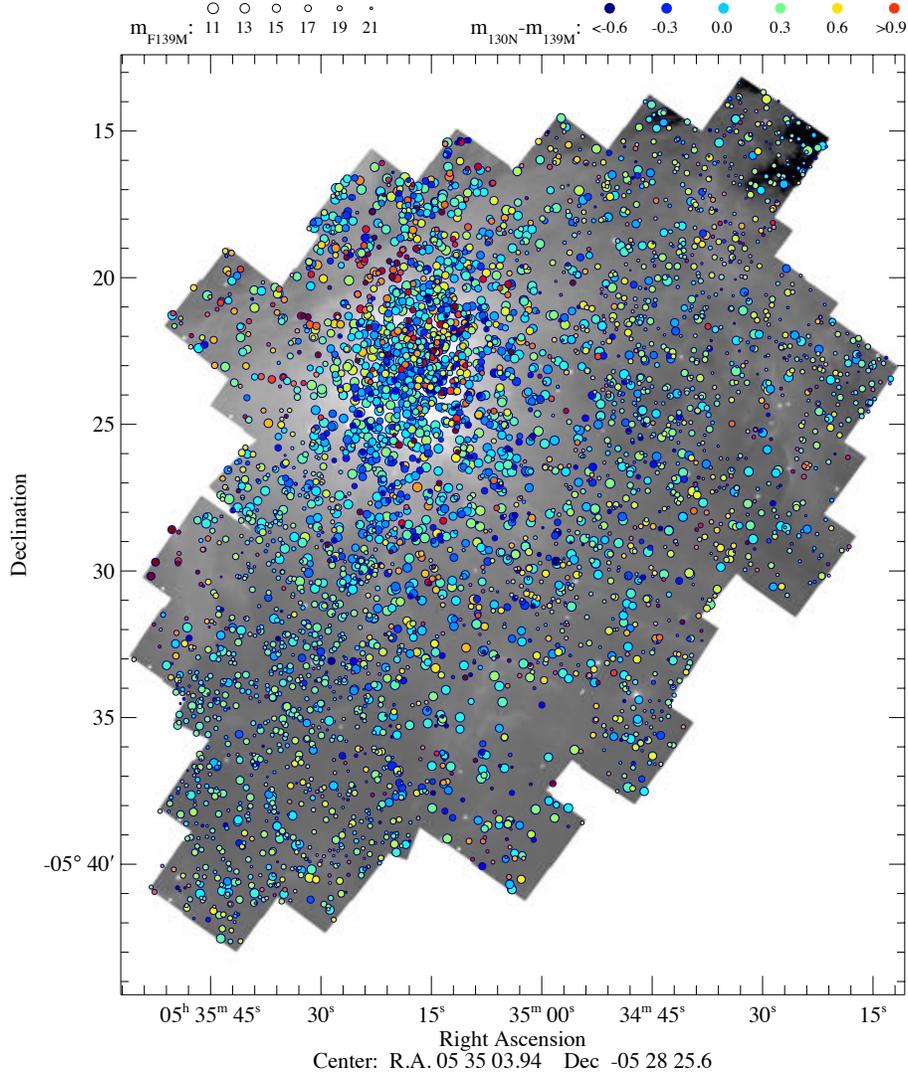}
\caption{
Survey area with superimposed the stars of our catalog. The symbol size scales with the magnitudes in the F139 filter while the color encodes the $m_{130}-m_{139}$ color index, as indicated by the legend at the top of the figure. \label{Fig:Stars in the Field}}
\end{center}
\end{figure*}

In Fig.~\ref{Fig:3plots} we present the calibrated magnitudes versus their corresponding errors for our two filters. Our $5\sigma$ sensitivity limits ($dm\simeq  0.2$) 
are at about $m\simeq 21.5$, in both filters, with a larger scatter in the narrower F130N filter. At the faintest flux levels, large errors ($dm\gtrsim 0.35$) in the F130N filter are compatible with non-detection.
The plot at the bottom shows the uncertainties in $(m_{139}-m_{130})$ color index, typically of the order of a few percent. 
In Fig.~\ref{Fig:Stars in the Field} we show the spatial distribution of our sources with symbols coded according to their brightness and color. The blue sources show H$_2$O in absorption and their distribution is generally centered around the Trapezium, { consistent with this being a spectral signature characteristic of the cluster members, in the range of massess probed by our survey}. The cyan-green sources do not show H$_2$O in absorption and are more uniformly distributed, with a surface density that tends to increase at the edges of the region. The red sources having high  $(m_{139}-m_{130})$ color appearing clustered e.g. to the north-east of the Trapezium, in correspondence of the dark filament that delimits the Orion Nebula, compatible with highly reddened objects.


\subsection{The photometric catalog}
Our final catalog lists 4,504 well measured sources; of these, 137 appear saturated in the F130N filter  and 221 in the broader $139M$ filter. Their brightness corresponds to sources with mass $\gtrsim 0.5$~\Msun, well above the mass range of interest; we include them in the catalog to facilitate cross-matching. 
The full list is presented in Table~\ref{tbl:catalog}, providing, in this order: 1) entry number, from 1 to 4504; 2-3 Right Ascension and Declination (J2000); 4-5) ($x$, $y$) pixels in the final mosaic images, Drizzled to a pixel scale of 0.1285\arcsec/pixel; 6-7) magnitude and uncertainty in the F130N filter; 8-9) magnitude and uncertainty in the F139M filter; 10-11) $(m_{130}-m_{139})$ color index, and uncertainty; 12) number of detections for each filter. 

\begin{deluxetable*}{cccccccccccc}
\tabletypesize{\scriptsize}
\tablecaption{WFC3-IR photometry of ONC sources\label{tbl:catalog}}
\tablewidth{0pt}
\tablehead{
\colhead{ID}\vspace{-0.2cm} & 
 \colhead{R.A.} & 
  \colhead{Dec.} & 
   \colhead{x} & 
    \colhead{y} & 
     \colhead{$m_{130}$} & 
      \colhead{$dm_{130}$} & 
       \colhead{$m_{139}$} & 
        \colhead{$dm_{139}$} &
         \colhead{$color_{130-139}$} &
          \colhead{$dcolor_{130-m_139}$)} &
           \colhead{detections}         \\
\colhead{\#} & 
 \colhead{(J2000)} & 
  \colhead{(J2000)} & 
   \colhead{(pixel)} & 
    \colhead{(pixel)} & 
     \colhead{(mag.)} & 
      \colhead{(mag.)} &
       \colhead{(mag.)} & 
        \colhead{(mag.)} &
         \colhead{(mag.)} &
          \colhead{(mag.)} &
           \colhead{\#} 
}
\startdata
   1 & 83.626381 & -5.457797 & 10112.94 &  7947.76 & $<9.8$   &  0.02 & $<9.78$  & 0.20&  -0.08 &  0.12&  3 \\
   2 & 83.608998 & -5.441777 & 10598.81 &  8397.33 & $<10.91$ &  0.02 & $<10.67$ & 0.03&  0.24  &  0.02&  4 \\
   3 & 83.616348 & -5.442953 & 10393.41 &  8364.36 &   11.72 &  0.01 &   11.53 & 0.02&  0.2   &  0.01&  2 \\
   4 & 83.632944 & -5.443575 &  9929.65 &  8347.02 &   12.15 &  0.03 &   12.13 & 0.01&  0.03  &  0.02&  2 \\
   5 & 83.617540 & -5.444363 & 10360.07 &  8324.79 &   13.38 &  0.01 &   13.43 & 0.02&  -0.05 &  0.02&  2 \\
   ...\
\enddata
\tablecomments{This table is published in its entirety in the electronic edition of the {\it Astrophysical Journal}.  A portion is  shown here for guidance regarding its form and content.}
\end{deluxetable*}

\section{Completeness analysis\label{Sec:Completeness}}
Artificial-star tests are a standard procedure to assess the level of completeness and precision of a photometric analysis. We performed our 
analysis using the capabilities implemented in the DOLPHOT package.
The first step is to create a list of fake stars, and we generated 16,000 artificial stars. The ($x ,y$) coordinates and the $m_{130}$ magnitudes were randomly generated with uniform density distribution in the range 16-24 mag, while the $m_{139}$ magnitudes were randomly generated from the cumulative distribution of the colors of the detected sources, for each magnitude bin. In this way the input magnitudes in the two filters reproduce the same range of colors covered by the data. 


We then inserted sequentially on each visit the artificial stars, checking how many times DOLPHOT would retrieve them. Following DOLPHOT prescriptions, 
we adopted the 
same parameter files used for the real detections, with the following additional parameters appropriate for the artificial star test: FakeMatch$=3.0$ (maximum allowable distance in pixels between input and recovered artificial star), FakePSF$=1.5$ (approximate FWHM of the artificial star in pixels), FakeStarPSF$=0$ (in order to use the PSF residuals from initial photometry run), FakePad$=0$ (artificial stars will be considered only if their center falls within any of the images), and RandomFake$=1$ (add Poisson noise to artificial stars).

\begin{figure*}[ht]       
 \includegraphics[width=\textwidth,trim={1.3cm 0.05cm 1.5cm 0.05cm},clip]{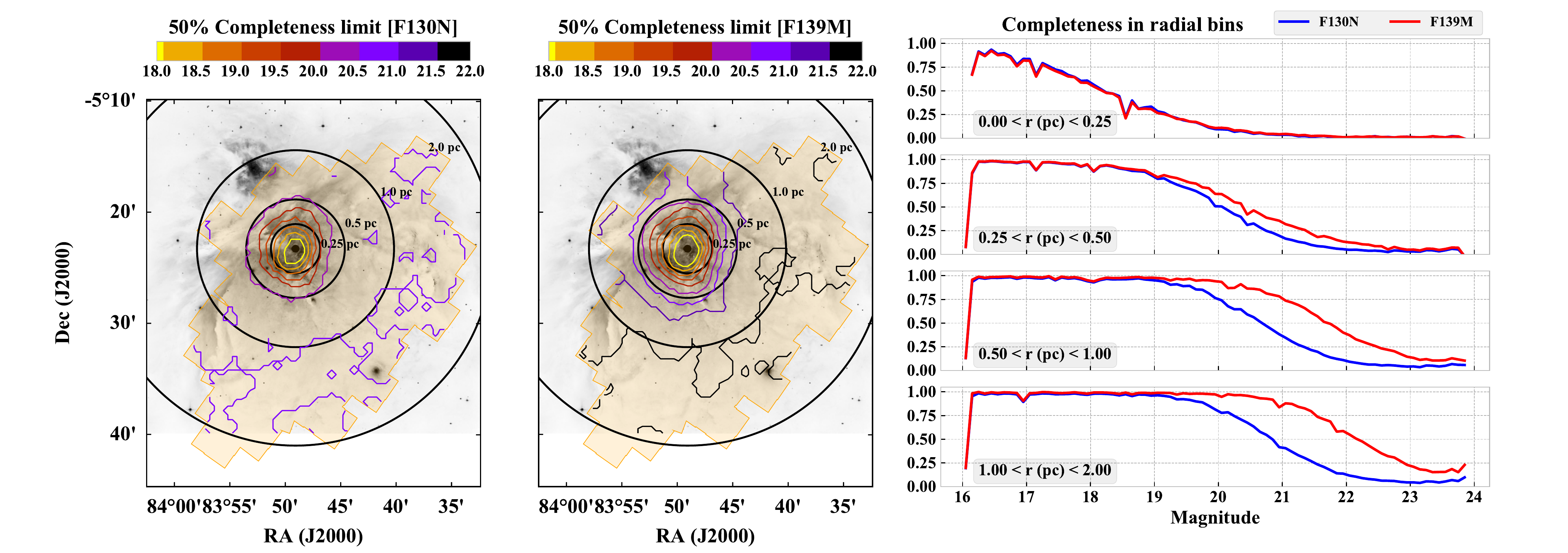}
 \caption{Spatial dependence of the completeness of our survey. The left and center maps are relative to the F130N and F139M filter, respectively.   The background image in gray-scale is an optical image of the region. Overlaid in sepia is the footprint of our near infrared observations. The color contours show the 50\% completeness limit in each band as a function of position in the sky. Their clustering at the center illustrates the rapid increase of completeness as one moves away from the bright cluster core. The black concentric circles, centered on (R.A., Dec.)$_{J2000}$  = (05:35:16.26, -05:23:16.4), correspond to 0.25, 0.5, 1, and 2 pc from the cluster center. 
These radii are used to define the radial bins of the right panel, showing how the average completeness, plotted as a function of the input magnitude, increases moving away from the center. 
 \label{fig:complcombo}}
\end{figure*}

Finally we applied to the recovered artificial stars the
same selection criteria used to automatically
discard spurious detections in our source catalog. 
In particular, we rejected stars where the difference between the input and output magnitudes of the recovered artificial stars was higher than 0.75 magnitudes in each filter.
The final result was a series of tables showing, for each visit, the fraction of recovered artificial stars per bin of magnitude, i.e., a spatial map of the photometric completeness at each magnitude level. 
In general, completeness is dominated by the nebular background, as illustrated in
Figure~\ref{fig:complcombo}: in the central region ($r<0.25$~pc angular distance from $\theta^1$~Ori-C, or about 3\% of our survey field), at the peak of the brightness of the nebular background, 50\% completeness 
is reached in both filters at $m\simeq18$. Moving away from the center, at projected distances larger than 1~pc, 50\% completeness is reached  $m_{130}\simeq 21$ and $m_{139}\simeq22$. Assuming the BT-Settl 1~Myr isochrone for the F139M filter, the average foreground extinction within our footprint \citep[$A_V \sim 2.2$ mag,][]{2011AA...533A..38S}, and a distance of 403~pc \citep[][see Appendix]{2019ApJ...870...32K} we conclude that our observations are 50\% complete down to about 10 \mjup\ in the very central region and down to 2.5 \mjup~ over 97\% of our survey field.  

\section{Results\label{Sec:Results}}
\subsection{Color-Magnitude Diagram}
\begin{figure}[htb!]
\begin{center}
\includegraphics[width=45 mm,angle=0,trim={2.0cm 0.25cm 2.8cm 0.25cm}]{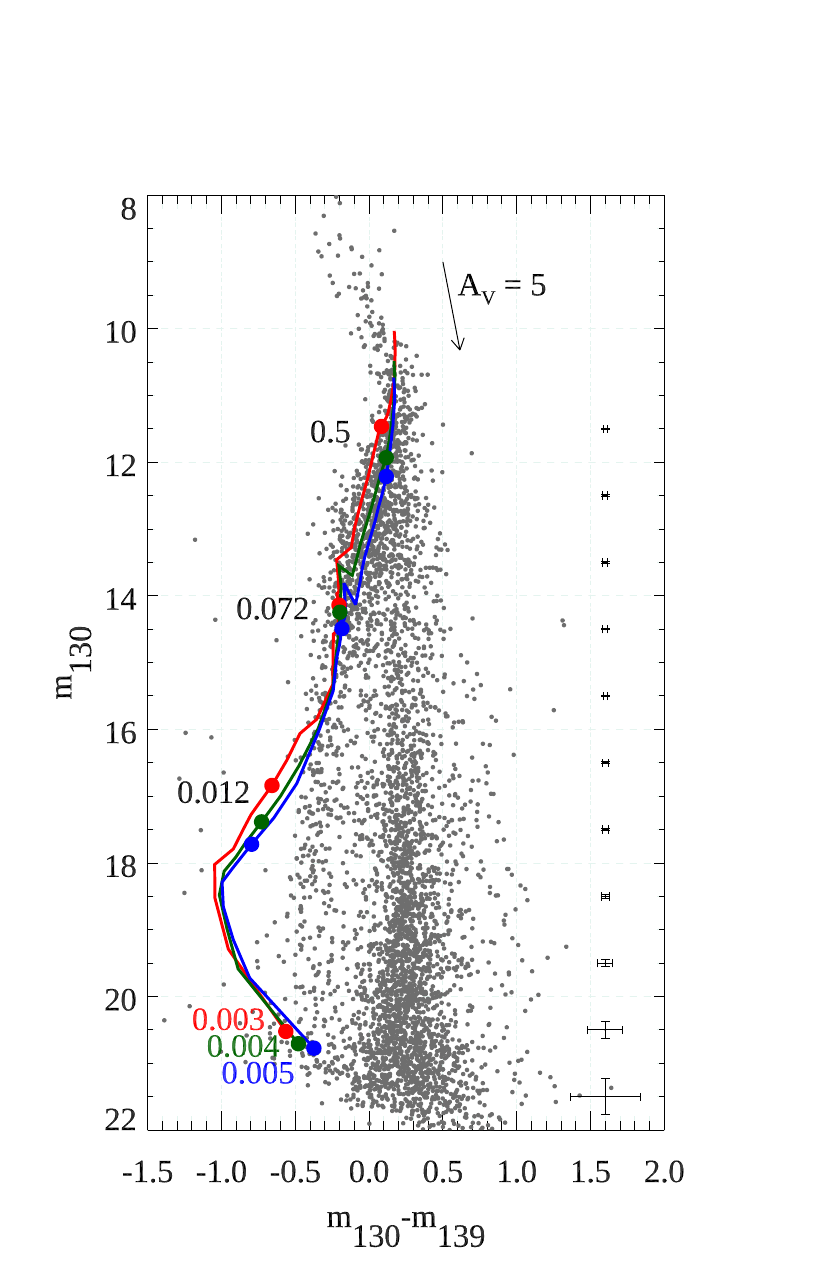}
\caption{Color-magnitude diagram for all sources detected in the F130N and F139M filters. The average photometric errors as a function of brightness are indicated on the right. The arrow indicates the reddening vector for $A_V=5$ assuming the \cite{1989ApJ...345..245C} 
reddening law for $R_V=3.1$. The three solid lines, top to bottom, represent the BT-Settl isochrones at 1~Myr (red), 2~Myr (green), and 3~Myr (blue). The filled dots on each isochrone correspond, top to bottom, to $M=0.5, 0.072, 0.012~\Msun$ and $0.003, 0.004$ and $0.005\Msun$, for the 1, 2, and 3~Myr isochrones, respectively. 
\label{Fig:BasicCMD}}
\end{center}
\end{figure}

In Figure~\ref{Fig:BasicCMD} we present the color-magnitude diagram for all sources ($n=450$). The data distribution shows a number of well defined features.  One can immediately recognize two main high-density populations: the clump at $m_{130}\sim12-14$ and the redder sequence in the lower half of the diagram becoming increasingly denser as the magnitudes increase. The former is dominated by the main stellar population of the ONC, with masses around the peak of the Initial Mass Function, in the 0.2-0.4~$\Msun$ range; the latter is the main background population seen through the backdrop of the Orion Molecular Cloud,  
most dense at faint magnitudes but with a tail towards brighter magnitudes of positive color.  The bright, cluster member sequence, also extends to fainter magnitudes but becomes increasingly displaced to negative index values. The lower half of the diagram at $m_{130}\lesssim 14$ thus shows two distinct populations: at the left of the main background population, one can recognize a second well-separated locus of sources that departs from the upper clump at $m_{130}\simeq 14$ and extends to the detection  limit at $m_{130}\sim21.5$, with increasingly bluer (negative) $(m_{130}-m_{139})$ color index. Negative color means that the 1.4~$\mu$m H$_2$O band is in absorption: we identify this sequence as the locus of the brown dwarf population of the ONC
and we count 742 sources with negative  $(m_{130}-m_{139})$ color index.

At the very top of the diagram ($m_{130}\lesssim 11$), the brightest sources also display increasingly bluer colors: this is an artifact due to saturation, that in the broader F139M filter occurs at lower flux levels than in the F130N filter. The $m_{130}$ magnitudes of these sources may still provide useful information and therefore they have been kept in our catalog.

In Figure~\ref{Fig:BasicCMD} we also plot the 1, 2 and 3~Myr isochrones
taken from the BT-Settl model { presented in Appendix A}, scaled to a distance { $d=403$~pc}, together with the points corresponding to the hydrogen and deuterium burning masses, assumed at $M=0.072~\Msun$ and $M=0.012~\Msun$, respectively, for solar metallicity \citep{2000ARA&A..38..337C}. 
The three curves are in good agreement with the data at $m_{130}\lesssim 15$, corresponding to about $0.045~\Msun$. Below this limit, however, there is a significant discrepancy, with the models predicting increasingly bluer colors, i.e., a deeper H$_2$O absorption feature down to $m_{130}\simeq18$ (corresponding to about $0.01~\Msun$) and then a return to redder colors down to our sensitivity limit $m_{130}\simeq21.5$ 
(that would correspond to $0.005-0.003~\Msun$ with about 2 visual magnitudes of reddening, depending on the assumed isochrone). For very-low masses, therefore, it is not possible to directly use the theoretical predictions for our bandpasses in order to derive the main stellar parameters. In Section~\ref{Sec:Isochrone} we will present an empirical recalibration of the BT-Settle 1Myr isochrone for $\teff\sim2500$~K in our two filters, anchored to the HST/ACS visible broad-band F850LP magnitudes and any prior information on membership and extinction. This requires an analysis of contamination, presented in the next three sections.

\subsection{Galactic Contamination}
\label{sec:galcont}

We perform a semi-quantitative assessment of galactic contamination using the Besan\c{c}on model \citep{2003A&A...409..523R0} of the Milky Way.
The tunable parameters of the model were set as: distance interval from 0 to 50~kpc, to capture the full Milky Way disk, thick disk and halo; field centered at Galactic coordinates ($l=209^{\circ}.1$, $b=-19^{\circ}.5$) with an area of 0.135 square degrees,
corresponding to the area of our survey;  diffuse extinction was left to nominal value (0.7~mag/kpc), as well as the range of stellar absolute magnitudes ($-7<20$) and spectral types (from O to OH/IR).

We additionally considered the extinction from the Orion Molecular Cloud (OMC). For all sources in the Besan\c{c}on model with distance greater than the distance to the ONC of about 400~pc, we used the extinction map from \cite{2011AA...533A..38S} to randomly assign a value of the total extinction $A_{V}$ peculiar to the Orion Nebula, to be added to the values adopted by the Galactic model. 

To derive magnitudes in our passbands, we used the noiseless apparent $V$ magnitudes from the Besan\c{c}on model and their diffuse $A_{V}$ to obtain the absolute $V$ magnitudes.
We then use \teff, $\log g$ and [Fe/H] from the model to assign a spectral energy distribution to the stars, using the \texttt{stsynphot} package\footnote{Available at \url{https://stsynphot.readthedocs.io/en/latest/}}. 
For the spectral energy distributions, we adopted the Phoenix grid of models available in \texttt{stsynphot} \citep{2013A&A...553A...6H}.
The spectra were normalized in flux using the absolute $V$ magnitude, then the total extinction (diffuse plus OMC, or diffuse only if $d <403$~pc) was applied to the spectrum. Finally, the magnitudes in F130N and F139M were obtained by integration under the respective throughput curves, using the appropriate zero-points.
Our artificial star tests was then used to account for photometric errors using the spatially closest artificial stars, with input magnitudes within 0.1 mag from the Besan\c{c}on model stars to allow for the usual underestimate of the photometric error in artificial star tests.

The results are shown in Figure~\ref{Fig:Contamination}. The CMD shows that a few foreground main sequence  dwarfs (purple dots) do, in fact, lie within the locus of cluster members. 
The rightmost panel of the figure, however, shows that their number is disproportionately low with respect to the bona-fide members. 
Most of the background contaminants (blue dots) are very well separated from the sequence of low-mass cluster members, another demonstration of the effectiveness of our photometric water index.
The histogram on the right panel also shows how the noise model we adopt (our artificial stars tests) is very effective in quantitatively predict how many contaminants should be observable. The observed and simulated luminosity functions do match very well down to our detection limit. 

\subsection{Extragalactic Contamination}
\label{sec:excont}
\begin{figure*}[htb!]
\begin{center}
\includegraphics[width=.9\textwidth, trim={0cm 1.cm 0cm 0.cm}]{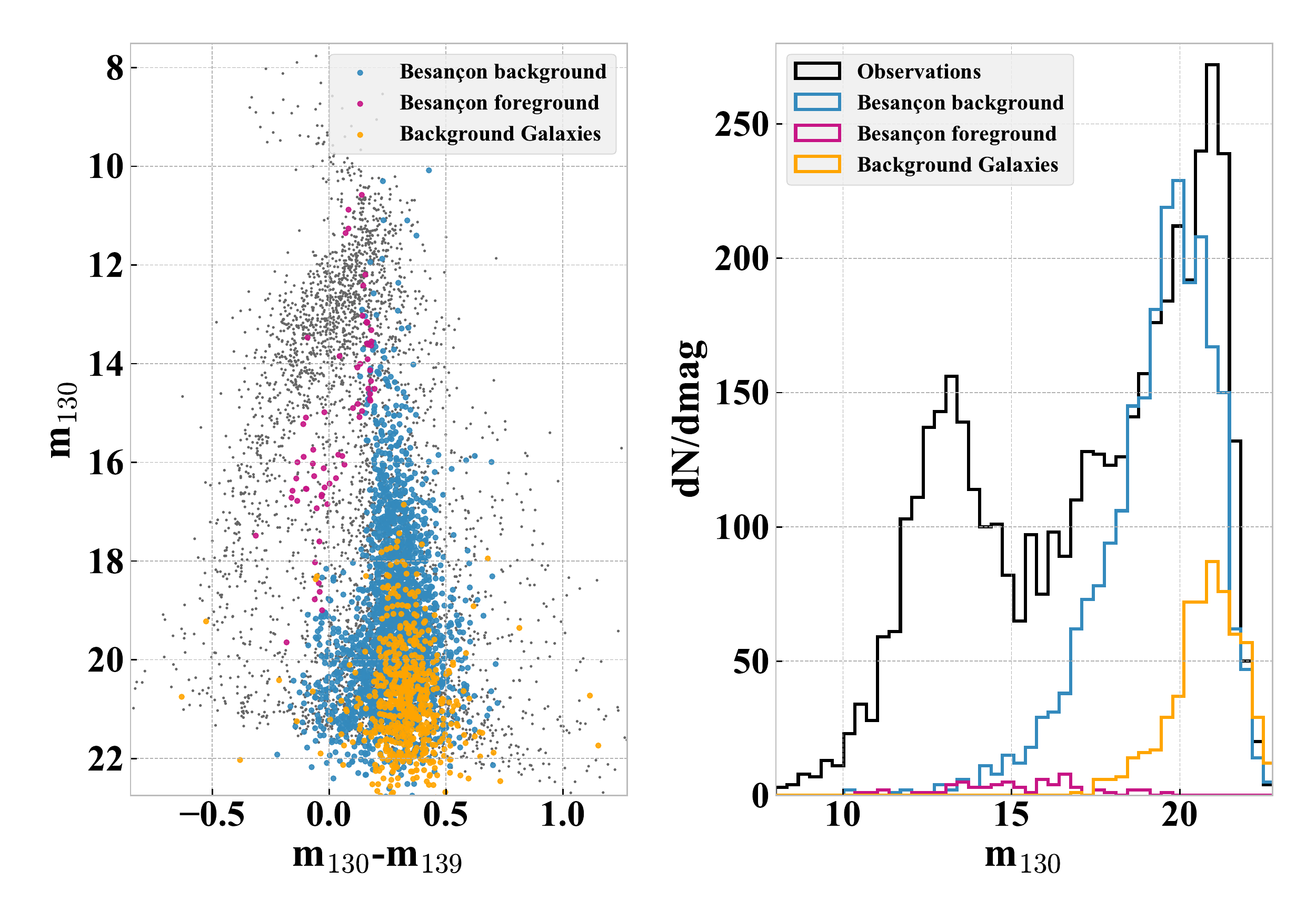}
\caption{Contamination from  stars, both foreground and background, and galaxies. Left: locus of the contaminants on the color-magnitude diagram; right: number distribution of the estimated population of contaminants vs. data. \label{Fig:Contamination}}
\end{center}
\end{figure*}

We also assess the amount of contamination to be expected from extragalactic sources. To do so we adopt a very similar procedure to that described in Section~\ref{sec:galcont} for the treatment of extinction and noise.
The spectra, however, are derived from best-fits to the multi-color photometry of all the sources available in the CANDELS fields catalogs \citep{2011ApJS..197...36K}, using the procedure described in \cite{2016ApJ...832...79P}.
The 5 CANDELS fields provide a sampling  of the expected cosmic variance  between randomly selected fields. We normalize the numbers of observed galaxies using our survey area and the areas of each CANDELS field; then we pick individual objects from the 5 fields at random and use their best fit spectrum to synthetically determine the photometry in our bandpasses. The results are shown in Figure~\ref{Fig:Contamination} as yellow dots.

Given that the best-fit template spectra used to fit the CANDELS HST photometry allow for a dependency on redshift and star formation history, among other quantities, for some galaxies in the sample there can be strong emission lines in either one of the F130N and F139M bandpasses. Galaxies with strong lines in F130N will appear bluer than $m_{130}-m_{139} = 0$ in the figure; vice-versa, the reddest objects may instead present emission lines in F139M.
Most of the extragalactic sources, however, are generally undetectable at the depths of our observations; for clarity, in the figure we do not show those falling below our sensitivity limits.
The large majority of observable extragalactic contaminants are relatively nearby galaxies dominated by stellar continuum emission in our passbands, and therefore they end up overlapping with the Galactic background stars in the CMD. Note that their number is far lower than background stars, in the magnitude ranged probed by our observations; these objects may be resolved by future high-spatial resolution observations.

{
\subsection{Contamination from Orion foreground population}
\cite{2014A&A...564A..29B} have found evidence for a foreground population in front of the Orion A cloud, deriving a distance of 380 pc and an age in the range 5-10~Myr. It turns out that for our sample this population represents a negligible source of contamination. Out the 605,020 sources contained in the full \cite{2014A&A...564A..29B} catalog, 1573 fall within the boundary of our survey. After cross-matching with our catalog, we find 24 sources having both membership $>99.5$ \citep[the criterion used by][to select foreground candidate members]{2014A&A...564A..29B} and color $m_{130}-m_{139}<0$. For 22 out of these 24 possible contaminants, the averages of the J and H-band magnitudes are in the range 12-14.5; this range (see our CMD) corresponds to the main stellar population of the ONC, where we count 670 sources with highest membership probability (Bayes Factor=2, see Section~\ref{Sec:BayesFactor}). The contaminant fraction is therefore of the order of 3\%; the remaining 2 sources of \cite{2014A&A...564A..29B} have fainter magnitudes and may contaminate our substellar population, traced down to planetary masses. But also in this case their number represents a  negligible fraction. If the Initial mass function maintains a negative slope going to even lower masses, it is safe to assume that the number of foreground contaminants remains extremely low also beyond the sensitivity limit probed by \cite{2014A&A...564A..29B}.
}

In conclusion, after considering all possible contamination from galactic, extragalactic and foreground sources, our catalog of bona-fide ONC members remains nicely uncontaminated, leading us to conclude that the large majority of sources with $m_{130} \gtrsim 14$ mag and $(m_{130}-m_{139}) < 0$~mag are likely to be ONC very-low mass members.  { In Section~\ref{Sec:BayesFactor} we will estimate the membership probability of each individual candidate on the basis of a Bayesian model of the cluster. This will require the assumption of a cluster isochrone for our two filters, as detailed in the following sections.}

\subsection{Comparison with previous spectral classifications \label{Sec:Comparison}}
To verify the accuracy of our photometric index, we have collected the spectral types presented by \cite{2012ApJ...748...14D}, complemented by the smaller catalogs \citep{2004ApJ...610.1045S, 2007MNRAS.381.1077R} of brown dwarf candidates,
all converted to \teff\ following \cite{2003ApJ...593.1093L}. 
Discussing their more recent compilation of spectroscopically determined spectral types,  \cite{2013AJ....146...85H} point out that for early-type M stars some systematic discrepancy may be presented between their spectral types and those determined by \cite{2012ApJ...748...14D} using narrow-band photometry in the TiO bands, whereas for mid- to late-M-type stars, the range probed by our survey,
narrow-band photometric methods are no less accurate on average than spectroscopic methods.

\begin{figure}
\begin{center}
\includegraphics[width=50 mm,angle=0,trim={2.0cm 0.25cm 1.8cm 1.8cm}]{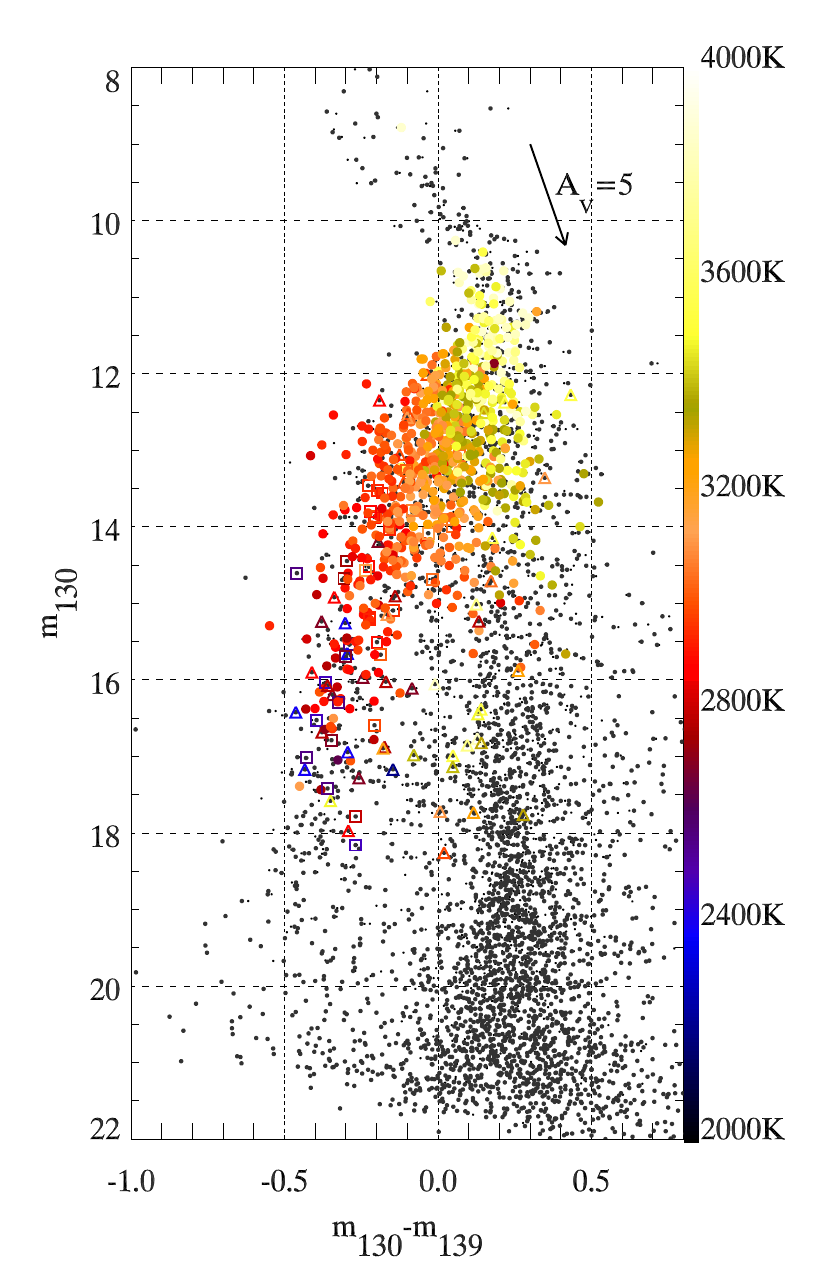}
\caption{Color-magnitude diagram highlighting sources that are known cluster members with determined effective temperature. The colors are coded according to  the temperature scale presented on the right axis. The symbols label the different source catalogs: circles for \cite{2012ApJ...748...14D}; triangles for \cite{2004ApJ...610.1045S}; squares for \cite{2007MNRAS.381.1077R}. 
\label{Fig:CMD_colorcoded}}
\end{center}
\end{figure}

Using the published \teff\ values, we can color code the sources with known spectral type on the CMD diagram. Figure~\ref{Fig:CMD_colorcoded} shows how \teff\ changes across the Color-Magnitude Diagram, with the colored bands tracing the locus of the Orion Nebula Cluster. The large majority of points (circles) comes from \cite{2012ApJ...748...14D}, whereas the spectroscopic surveys of \cite{2004ApJ...610.1045S} (triangles) and \cite{2007MNRAS.381.1077R} (squares) provide the effective temperature for the lowest mass sources (blue colors). There is both a substantial vertical spread and a clear correlation with the strength of the H$_2$O index when \teff$\lesssim 3200$~K.

Figure~\ref{Fig:CMD_colorcoded} shows how the large majority of sources belonging to the galactic population remains too faint for spectral classification due to the large extinction from the Orion Molecular Cloud. It also shows the gain in sensitivity and diagnostic power achieved by our HST observations over previous surveys, as we are able to trace the substellar population of the cluster well beyond the $m_{130}\simeq 17$-equivalent limit reached by spectroscopic surveys; the majority of ONC sources fainter than $m_{130}\sim15$ have no previously determined spectral type. 

The color spread in Figure~\ref{Fig:CMD_colorcoded} cannot be entirely explained as the result of a narrow isochrone spread by random values of extinction. { A similar plot for dereddened magnitudes (not shown here) would looks quite similar to Figure~\ref{Fig:CMD_colorcoded}, as expected 
since  the median extinction of the sources with known spectral type is $A_V\simeq2$ and about 90\% of the sources are known to have $A_V\lesssim5$ \citep{2012ApJ...748...14D}. }
Instead, the vertical spread must be regarded as another indication of an intrinsic luminosity spread, confirming what found by \cite{2011A&A...534A..83R} by analyzing HST photometry in visible bands, and inferred by \cite{2011MNRAS.418.1948J} through different age indicators.

\subsection{An H$_2$O spectral classification index}\label{sec:H2Oindex}
\begin{figure}[htb!]
\begin{center}
\includegraphics[width=90 mm,angle=0]{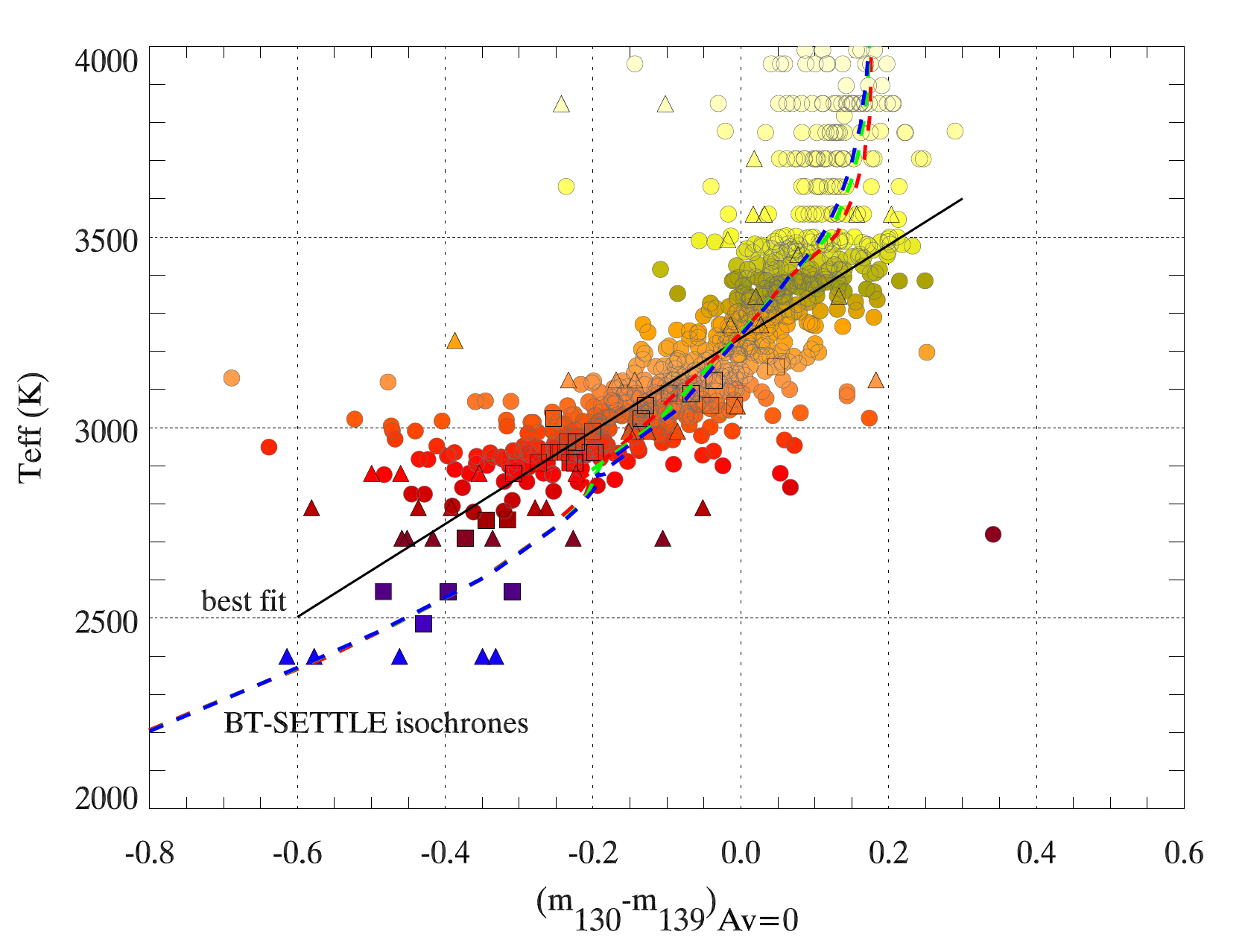}
\caption{Relation between the effective temperatures reported in the literature and our H$_2$O index. The symbols and color scale are the same of Figure~\ref{Fig:CMD_colorcoded}. The red, green and blue dashed lines, barely distinguishable, refer to the 1, 2 and Myr isochrone of the BT-Settl models. The solid line represent the best linear fit to the data (see text). \label{Fig:water_Av}}
\end{center}
\end{figure}

The sources with known spectral type allow us to derive a relation between \teff\ and the intrinsic $(m_{130}-m_{139})$ color index, i.e., a H$_2$O spectral index. In Figure~\ref{Fig:water_Av} we plot these two quantities, where the colors have been dereddened using the values of $A_V$ provided by \cite{2012ApJ...748...14D}, \cite{2004ApJ...610.1045S}, and \cite{2007MNRAS.381.1077R} and the following relations:
\begin{eqnarray}
A_{130}&=0.264 A_V,\label{eq:Alambda1}\\
A_{139}&=0.241 A_V \label{eq:Alambda2}
\end{eqnarray}
valid for the $R_V=3.1$ \cite{1989ApJ...345..245C} reddening law. These relations have been derived using stsynphot \citep{Laidler+08} to extract synthetic photometry of late type spectra, but using the  values of $A_V$/$A_\lambda$ at the effective wavelength of the filter provides nearly identical values as we are dealing with relatively narrow filters.

The plot shows that our H$_2$O spectral index starts to be sensitive to \teff\ below $\lesssim 3500$~K;
Given the scatter and low numbers at the bluest index values, we perform a basic outlier-resistant two-variable linear regression of all data with \teff $<3500$ (roughly corresponding to spectral type M2) and H$_2$O index $>-0.3$, deriving the following equation (solid line in Figure~\ref{Fig:water_Av})
\begin{equation}
T_{\rm eff}~({\text K})= 3235 + 1020 \times (m_{130}-m_{139}),
\end{equation}
with an average scatter around the best fit line of 98~K. We also plot for comparison the 
1-3~Myr isochrone from the BT-Settl models (see Appendix A) as a dashed line, that appear to systematically undestimate the effective temperature below $\teff\sim3000$~K.
As remarked by \cite{2012ApJ...748...14D}, the presence of accretion would create veiling of the spectral signatures, reducing the measured depth of the $1.4\mu$m absorption feature and therefore producing an H$_2$O index redder than the pure stellar isochrone; this implies that the discrepancy between the model and our observations cannot be reconciled by assuming some unaccounted contribution from extra accretion luminosity.  

\subsection{Stars with ACS detection in the F850LP filter }
Using the source catalog obtained from our 2005 
HST Treasury Program \citep{2013ApJS..207...10R}, we combine our near-IR WFC3 data with previous ACS photometry at visible wavelengths. The deepest optical data have been obtained with the F775W and, especially, in the long-pass filter F850LP. 

\begin{figure}[htb!]
\begin{center}
\includegraphics[width=90 mm,angle=0]{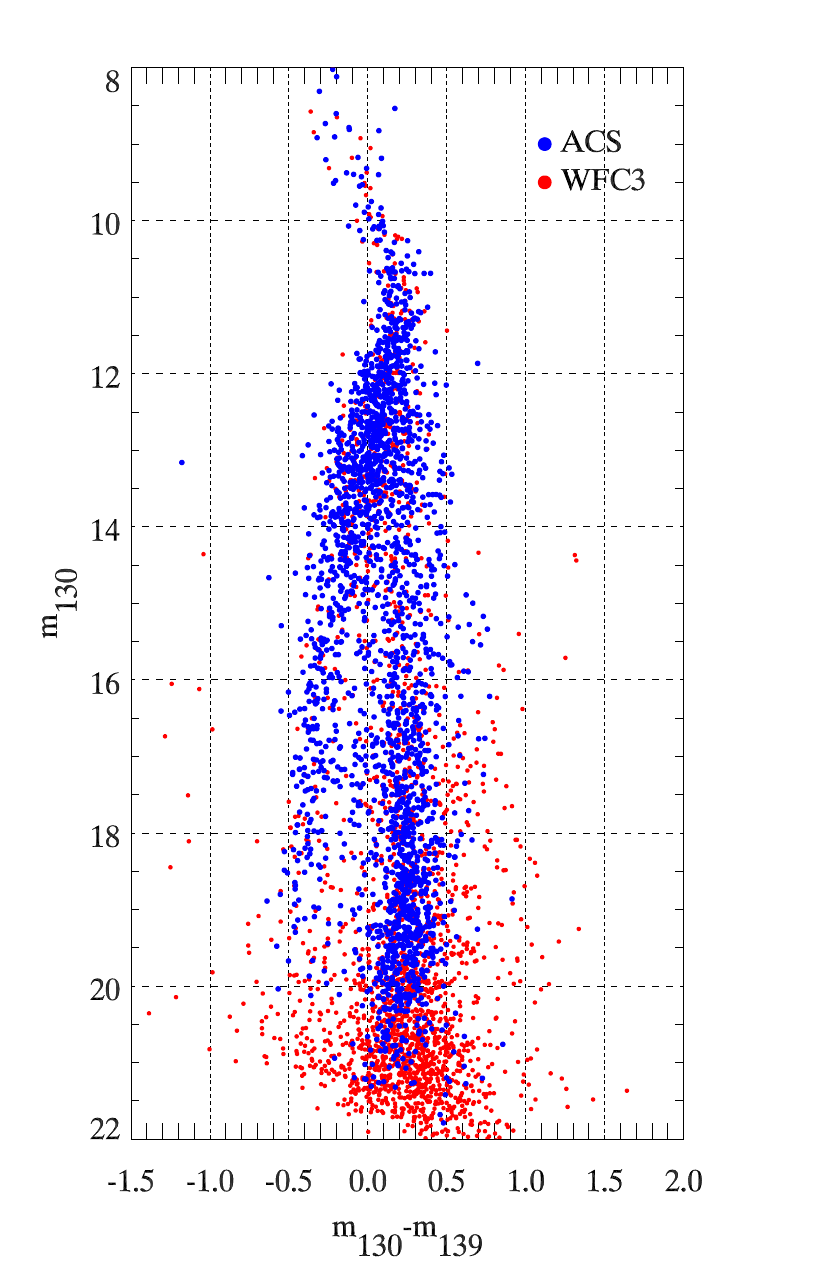}
\caption{WFC3-IR Color-Magnitude Diagram, with sources detected in the F850LP filter overplotted as blue dots.  \label{Fig:CMD+ACS}}
\end{center}
\end{figure}

{ Cross-matching the two catalogs by assuming a maximum distance of $0.2''$ between the celestial coordinates, we find  that 2754 ACS sources have been also detected in our near-IR bands. The combined photometry is presented the in Table~\ref{tbl:ACS+WFC3}. 
As shown in Figure~\ref{Fig:CMD+ACS}, the sources detected with both ACS and WFC3 are evenly distributed across our near-IR CMD. Overall, the depths of the two HST Treasury programs on the Orion nebula, GO-10246 and GO-13826 are comparable, meeting our original survey requirements.}


\begin{deluxetable*}{ccccccccccccc}
\tabletypesize{\scriptsize}
\tablewidth{0pt}




\tablecaption{Cross-matched ACS-VIS and WFC3-IR photometry of ONC sources
\label{tbl:ACS+WFC3}}


\tablehead{\colhead{index} & \colhead{R.A. (J2000)} & \colhead{Dec. (J2000)} & \colhead{$m_{775}^{*}$} & \colhead{$dm_{775}^{*}$} & \colhead{$m_{850}^{*}$} & \colhead{$dm_{850}^{*}$} & \colhead{$m_{130}$} & \colhead{$dm_{130}$} & \colhead{$m_{139}$} & \colhead{$dm_{139}$} & \colhead{$color_{130-139}$} & \colhead{$dcolor_{130-139}$} \\
\colhead{} & \colhead{degrees} & \colhead{degrees} & \colhead{mag.} & \colhead{mag.} & \colhead{mag.} & \colhead{mag.} & \colhead{mag.} & \colhead{mag.} & \colhead{mag.} & \colhead{mag.} & \colhead{mag.} & \colhead{mag.} } 

\startdata
   3 & 83.61634780 & -5.44295295 & 14.249 & 99.990 & 13.414 99.990   & 11.724 0.011  &  11.530 & 0.018  & 0.195 & 0.007\\
   4 & 83.63294382 & -5.44357471 & 14.434 & 99.990 & 13.567 & 99.990   & 12.146 0.030  &  12.132 & 0.012  & 0.031 & 0.021\\
   5 & 83.61754036 & -5.44436287 & 16.037  & 0.002 & 14.901  & 0.002   & 13.378 0.010 &   13.427 & 0.023 & -0.053 & 0.018\\
   7 & 83.62495811 & -5.42146638 & 15.662  & 0.002 & 14.705  & 0.002   & 13.228 0.013 &   13.309 & 0.014 & -0.093 & 0.016\\
...\\
\enddata

\tablenotetext{*}{In order to account for variability, Table~5 of \cite{2013ApJS..207...10R} reports the magnitudes of each individual detection. To compile this table we have taken the average magnitudes when multiple values have been reported. Following the same convention adopted by \cite{2013ApJS..207...10R}, a magnitude value equal to 99.99 indicates a value below the detection limit reported in the adjacent column for the magnitude error. Viceversa, a magnitude error equal to 99.99 refers to a saturated star brighter than the magnitude reported in the corresponding column for the source magnitude.}
\tablecomments{This table is published in its entirety in the electronic edition of the {\it Astrophysical Journal}.  A portion is  shown here for guidance regarding its form and content.}

\end{deluxetable*}


\begin{figure}[htb!]
\begin{center}
\includegraphics[width=85 mm,angle=0, trim={2.0cm 0.5cm 1cm 0.cm}]{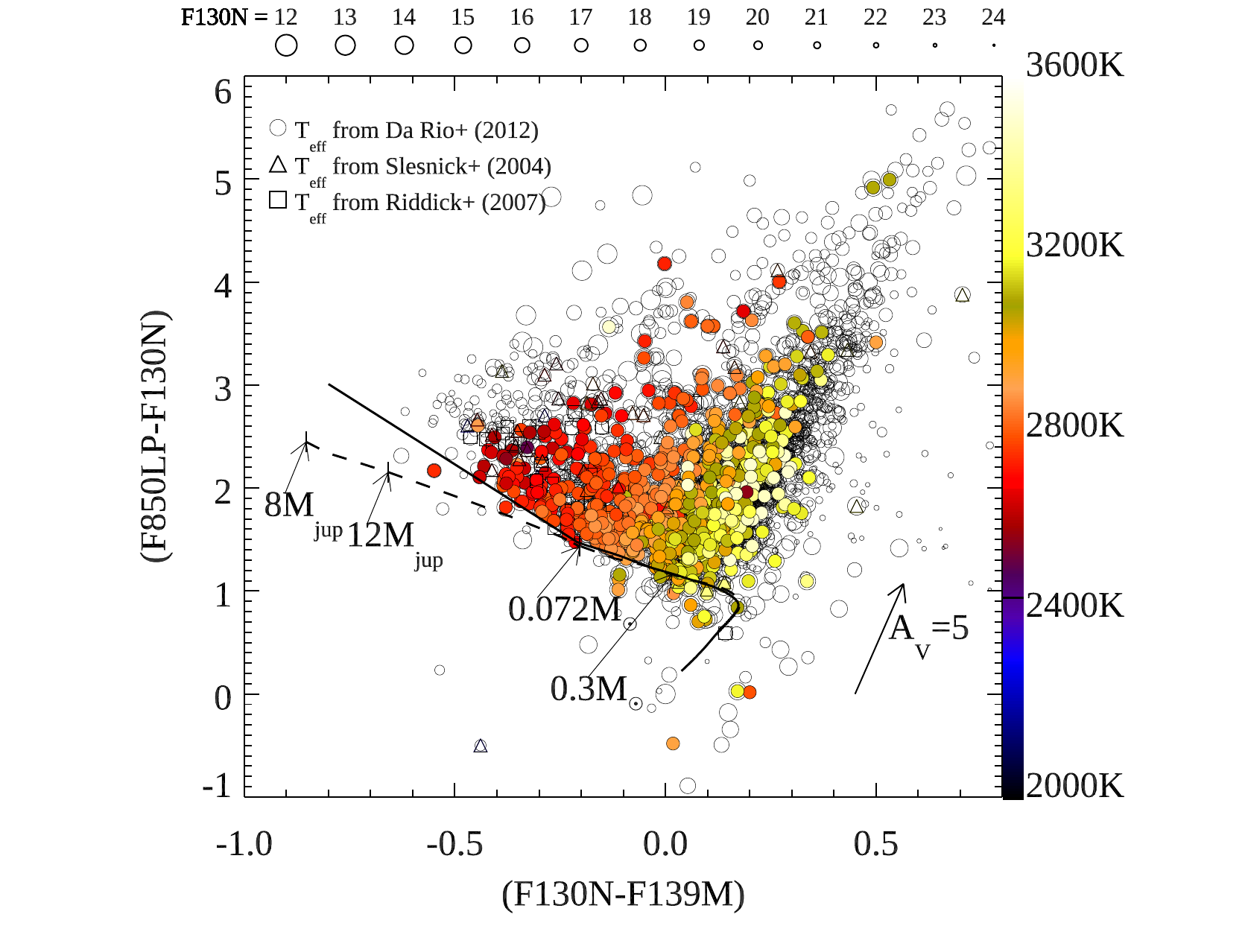}
\caption{$m_{850}-m_{130}$ vs. $m_{130}-m_{139}$ color-color diagram showing the  distribution of sources with known spectral type (color coded accrording to the temperature scale presented on the right axis) and unknown spectral types (gray open circles); the size of the circles is indicative of the source brightness in the F130N filter, according to the scale presented on the top axis. The dashed line indicates the locus of the BT-Settl 1~Myr isochrone. The solid line is our slighlty revised line better matching the envelope of the data points at low masses. \label{Fig:2CD}}
\end{center}
\end{figure}

With a third photometric band, we can create a distance independent $(m_{850}-m_{130})$ vs. $(m_{130}-m_{139})$ color-color diagram, shown in  Figure~\ref{Fig:2CD}, analogous to the $(753-I)$ vs. $(753-770)$ diagram presented by \cite{2012ApJ...748...14D} for the 6100~\AA ~ TiO band.
The plot shows again the decrease in \teff\ as one moves leftward to bluer color indexes. The spread along the reddening vector also decreases, consistent with the fact that fewer very-low-temperature and fainter objects have been identified and classified.
For spectral indexes $(m_{130}-m_{139}) \gtrsim 0.2$, the 1~Myr BT-Settl model isochrone (dashed line) nicely matches the lower envelope of the data distribution, i.e., the locus of the dereddened sources. Beyond this value, the model isochrone  predicts lower $m_{850}-m_{130}$ color index, another indication of the current limits of the models in our passbands. At the right side of the plot, many sources with high reddening have no spectral type: these are faint, highly reddened background stars. 
%

Following \cite{2012ApJ...748...14D}, we have slightly modified the isochrone to make sure that it matches more precisely the lower envelope of the data points. 
We match the theoretical isochrone above $(m_{130}-m_{139})=-0.2$, corresponding to the location of a 0.07~\Msun star in the 1~Myr uncorrected BT-Settl isochrone. For bluer colors and smaller masses, our line departs sligthly from the model. The black solid line in Figure~\ref{Fig:2CD} is given by the equations:

\begin{equation}
m_{130} = A \times (m_{130}-m_{139}) + (m_{850}+B) 
\label{eq:2CD}
\end{equation}
where $A=2.7,~ B=-0.85$ when $(m_{130}-m_{139})>-0.2$ and $A=1.38,~ B=-1.13$ below this value. 
This equation sets a first constraint on the source colors that we will use in the next section to refine the cluster isochrone in our CMD. 

Ideally, it should be possible to derive the extinction $A_V$ of each source in this diagram by projecting the data point on the isochrone. However, since the ACS F850LP data and the present WFC3 observations were taken at different epochs, the results would be affected by stellar variability. According to \cite{2002A&A...396..513H}, essentially every ONC star brighter than $m_I=16$ is variable, with about 50\% having peak-to-peak variations $\sim 0.2$ magnitudes or more in the Cousin-I band. A change of 0.2 magnitudes in our near-IR filters would corresponds to an error in the estimated value of $A_V$ of 0.8 magnitudes. In general, simultaneous wide-band measures are needed to reliably estimate the extinction toward individual young stars, and consequently all derived stellar parameters like luminosity, radius, isochronal age and mass.
{
\subsection{Veiling \label{Sec:Veiling}}
Veiling from mass accretion may potentially affect our color index. To assess its relevance, we can refer  to \cite{2005AN....326..891M} who show that the mass accretion rate in young ultra low-mass objects down to nearly the deuterium-burning limit appears on average to decrease steeply with mass, e.g. $\dot{M}\propto M^{-2}$.  Below the H-burning limit, $\dot{M}$ values are typically in the range $10^{-11}-10^{-12}$\Msun yr$^{-1}$. If we take as a reference a 0.07 \Msun~ star, and compare the accretion luminosity vs bolometric luminosity from the BT-Settl model for a 1~Myr isochrone, we have accretion luminosities of the order of 10$^{-4}$ the bolometric luminosity of the brown dwarf, extremely small. Moreover, the peak of $\simeq2500$~K stellar spectra falls in the near-IR, where the $H_2O$ feature is prominently observed; viceversa, accretion continuum is generally approximated as a $\sim10,000$~K blackbody plus emission lines. Both factors, total luminosity and spectral distribution, allow us to conclude that the contribution of accretion veiling in our data is expected to be negligible. 
}
\subsection{Isochrone \label{Sec:Isochrone}}


A direct comparison of our dataset with the theoretical isochrone, as presented in Figures~\ref{Fig:BasicCMD} and \ref{Fig:2CD}, clearly shows that a BT-Settl model predicting the $H_2O$ color down to planetary masses does not agree with the observations below \teff $\sim2500$~K. The causes of the discrepancy in the F139M band may be related to the line lists and opacity tables used to predict the water absorption feature at low effective temperatures. In fact, the data we are presenting are the first to provide a sample large enough to unveil systematic differences between theory and observations in near-infrared bands not accessible from the ground. This type of dataset may thus help constraining theoretical predictions of the near-infrared spectra of young brown dwarfs and planetary mass objects. 

In order to provide an initial raw estimate of the physical parameters of our very-low luminosity sources, we apply an empirical correction to the model predictions in our IR bandpasses using the observational constraints inferred from both the near IR CMD (Figure~\ref{Fig:CMD+ACS} and the color-color diagram build combining ACS and IR data (Figure~\ref{Fig:2CD}.
Our goal is to shift the theoretical isochrone in our CMD toward an empirical locus 
that better describes the average location of the ONC stars. Our corrections modify the model predictions for our two IR bands, but not for the broad-band F850LP filter, and it is not arbitrary, being constrained by the relation between $(m_{850}$, $m_{130})$ and $m_{139}$ determined with the color-color diagram (Equation~\ref{eq:2CD}): our adjusted empirical IR isochrone, when plotted in the color-color diagram, must still lie along the lower envelope of the data points, i.e. the envelope corresponding to stars with $A_V = 0$. 
The models correctly predict the $m_{850}$ magnitudes, 
which we leave unchanged, while a correction is provided in both the $m_{139}$ magnitude and in the $(m_{130}-m_{139})$ color index. 
In the Appendix we describe in detail the process used to derive our empirical correction. 
Table~\ref{tab:BT-Settl} lists the corrections as a function of \teff\ along the isochrone. In this way the correction, which is derived using the 1~Myr isochrone, can be applied to other isochrones, assuming that the observed discrepancies are function of the model temperature only.


\subsection{Membership probability}\label{Sec:BayesFactor}
In order to quantify the probability of each source in our catalog to be either a cluster member or a background contaminant, we calculate the ratio of the marginal likelihood of our data given these two hypotheses. The results for an individual star, characterized by its color and magnitude measurements arranged in a 3-elements vector $\mathbf{d}_i$, are expressed in terms of the Bayes Factor, given by
\begin{equation}
    BF_i = \frac{p(\mathbf{d}_i|ONC)}{p(\mathbf{d}_i|\overline{ONC})}.
\end{equation}
A Bayes Factor $BF=1$ indicates equal probability of star being cluster or background source. The marginal likelihood for the membership hypothesis, $p(\mathbf{d}_i|ONC)$, is obtained by marginalizing over a model of the CMD for the ONC cluster, while the likelihood for the non-member case, $p(\mathbf{d}_i|\overline{ONC})$ is marginalized over a model of the CMD for background contaminants.
The procedure for marginalization follows the scheme by  \cite{2018AJ....156..196R}.

The cluster model adopted here is described in detail in the accompanying Paper~II 
on the ONC IMF, so here we only summarize the main points.
We describe the CMD of the ONC population using the 1 Myr
isochrone derived in the previous section. The mass distribution is described as a 3-part broken power law, similar to the \cite{2001MNRAS.322..231K} IMF, the parameters of which are determined in Paper~II.
To populate the cluster CMD model, we draw a parameter set from the IMF posterior and use such parameters, which include a normalization factor and a value for the binary fraction, to draw a given number of masses.
Use the 1 Myr isochrone we assign a color and a magnitude value to each mass. We also draw on-sky positions according to the 3D model of the ONC also described in Paper~II (a spherically symmetric model with a radial density profile declining like a power-law, a model that reproduces very well the observed radial counts profile).
Positions and intrinsic magnitudes, together with the reddening map by \cite{2011AA...533A..38S} and the artificial star tests, allow us to produce a realistic CMD for a model-ONC, which includes measurement errors and incompleteness.
The experiment is repeated by drawing multiple IMF parameter sets, thus marginalizing over the posterior distribution of IMF parameters, until 2 million stars are drawn.
The model probability is then derived by using a kernel-density estimate, in CMD space, which is normalized to unity over the entire CMD.
A similar approach is used for the background. In this case, the simulated CMD is derived from the union of the Besan\c{c}on model, for galactic contaminants, and an estimate of the extragalactic contaminants using the CANDELS fields (see Sections ~\ref{sec:galcont} and \ref{sec:excont}).
The cardinality of the $ONC$ and $\overline{ONC}$ CMD models priors are based on the number of cluster sources identified based on a simple color-cut argument, and the number of expected background sources from the Besan\c{c}on model and the area-scaled CANDELS counts.

\begin{figure*}[htb!]
\begin{center}
\includegraphics[width=180 mm,angle=0]{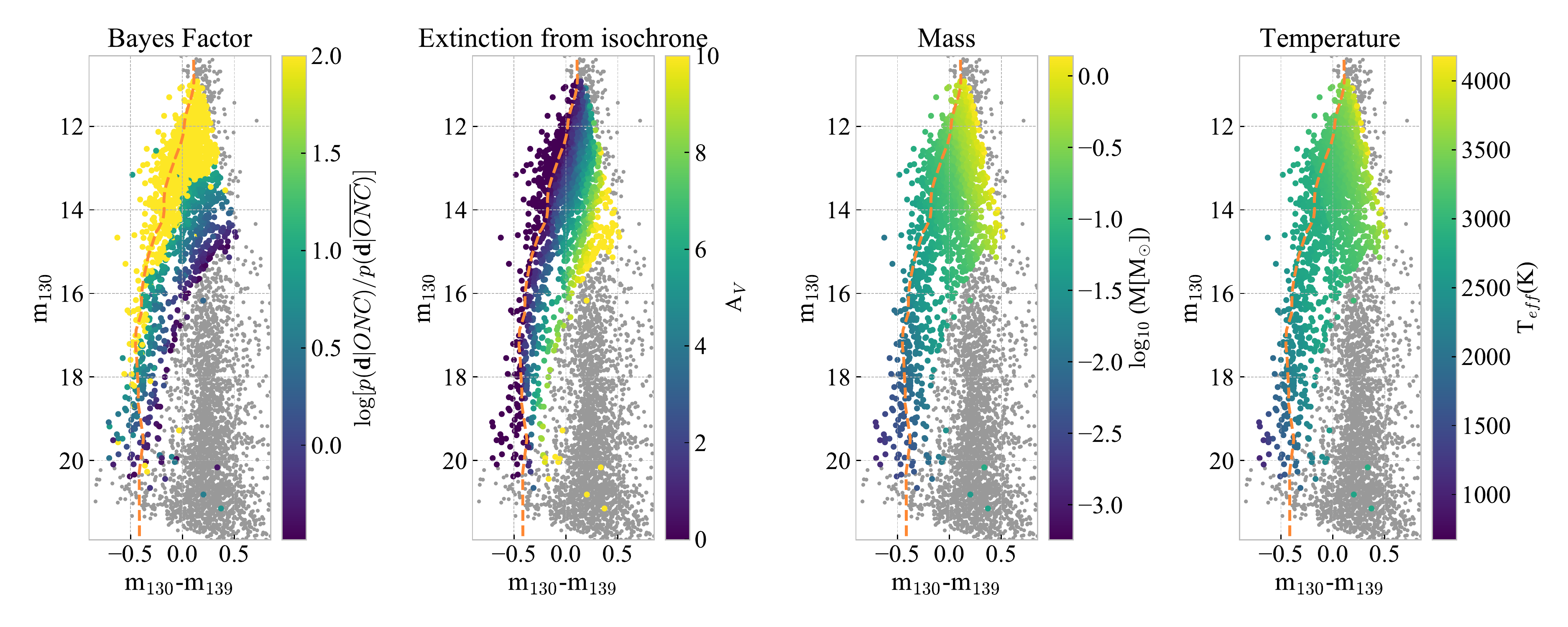}
\caption{
Gray dots at the faint end of the cluster population are due to the fact that, in our calculation of the Bayes Factor, we apply a completeness cut at the faint end. Thus, for this faint-end sample below our completeness threshold we do not compute membership probabilities. The dashed orange line represents our semi-empirical isochrone.
\label{Fig:AMTB}}
\end{center}
\end{figure*}

Our results are shown in Figure~\ref{Fig:AMTB} and are broadly in line with the more cursory color-based approach. Our threshold for declaring a source a ONC candidate, i.e. showing some evidence that it may not belong to the background population, corresponds to logarithms of the Bayes Factor $>-0.5$. Bona fide cluster members have Bayes Factor $>1$
Our approach based solely on the H$_2$O CMD position of the objects represents a good first order approach to the probability of ONC membership, specially on a global scale; it remains clear that for individual sources higher confidence can be reached when additional information (e.g., spectra or accurate proper motion) is available and taken into account.

\subsection{Estimate of the stellar parameters}
Having derived a bona-fide empirical isochrone in our filters and a measure of the membership probability, we can try to deredden our candidates in the CMD to derive a raw estimate of their effective temperature and mass. 
We expect significant uncertainties in our derived stellar parameters. To begin with, the spread in luminosity discussed in Section~\ref{Sec:Comparison} and other uncertainties result in a number of sources lying in the CMD above the empirical isochrone; their projection on the isochrone along the reddening vector requires unrealistic negative values for the extinction; we assign to them $A_V=0$. For sources below the isochrone, we determine their $A_V$ dereddening their photometry to the isochrone. As a result, the derived values of extinction can be quite different from those estimated through direct spectroscopic measurements. With these caveats, the results of our exercise are graphically shown in Figure~\ref{Fig:AMTB} and 
presented in tabular form in Table~\ref{Tab:Calalog_AMT_BR}. The list contains 1684 stars, having neglected all sources with $\log(BF)<-0.5$ 
having probability larger than 80\% of being contaminants, and sources saturated according to the magnitude limits reported in Sec~\ref{sec:photometry}.
The extinction diagram shows the expected nearly-vertical gradient along the reddening direction, with the largest $A_V\sim15$ values reached by the more massive stars in the upper part of the diagram. The masses diagram shows that masses span the entire range from sub-solar to a few Jupiter masses, while the temperatures range between about 4000~K and 1500~K.

\subsection{Low mass stars to brown dwarf ratio}
{ 
A parameterization often used in the literature to characterize the low-mass end of the IMF is the ratio $R$ between stars and brown dwarfs. For the ONC, \cite{2011A&A...534A..10A} investigated how $R$ changes with the radial distance from the cluster center exploiting a HST/NICMOS survey \citep{2013ApJS..207...10R} covering $\sim28\%$ of the region in J- and H-equivalent bands .
After correction for field star contamination, that they estimated at a 10\% level for brown dwarfs down to a mass limit of 30~\mjup, \cite{2011A&A...534A..10A} found a cluster averaged value $R=2.4$ with strong radial dependence, from $R=7.2\pm5.6$ in a inner $d=0.35-0.7$~pc 
annulus to  $R=1.1\pm0.8$ in the outer $d= 1.1-1.5$~pc annulus. If confirmed, a radial decrease of the stars to brown dwarfs ratio could suggest that mass segregation can be traced down to the subsolar masses, or that brown dwarfs are already dynamically spread at $t\sim1$~Myr, constraining the star formation history of the ONC. The reported uncertainties, however, are too large too draw conclusions. To put these $R$ values in a more general context, one can refer e.g. to Table 4 of \cite{2012ApJ...744....6S} that lists the $R$ values for 8 young clusters. Their values range from $R=2.3$ for NGC~1333, the object of their study, to $R=8.3$ for IC 348, the value reported by \cite{2008ApJ...683L.183A}. More recently, \cite{2019ApJ...881...79} found values in the range $R=2-5$ for 7 clusters.  A solid confirmation of a spread in the $R$ values for nearly coeval young clusters would also have major implications, pointing to variations of the Initial Mass Function at substellar masses, but also in this case uncertainties are large.

If we adopt the same mass bins used by \cite{2011A&A...534A..10A}, i.e.  0.08-1.0~M$_\odot$ for the stars and 0.03-0.08~M$_\odot$ for brown dwarfs, we  obtain $R$ values that strongly depend on the Bayes Factor, i.e. on our estimated membership probability. Figure~\ref{Fig:Rvsdistance} shows the dependence of our $R$ values vs. the Bayes Factor for the four radial zones of \cite{2011A&A...534A..10A}. In general, it is $R\simeq 6-7$ for Bayes Factor $\sim2$, our highest level of confidence. The dependence of $R$ on the Bayes Factor does not depend on the distance from the cluster center, with the possible exception of the outer region that shows a peak at Bayes Factor $\simeq 0$, a feature caused by a relative increase of stellar contaminants vs brown dwarf contaminants for that $R$ value.

Making a step further, in Figure~\ref{Fig:Rvsdistance_Planets} we present the corresponding $R$ ratio for brown dwarf (0.015-0.08~M$_\odot$) vs. planetary mass objects (0.05-0.015~M$_\odot$). 
The spread of $R$ ratios for for Bayes Factor = 2 is larger, with a peak at $R\simeq10$ for the $0.7<d<1.1$~pc region, but one has to take into account the low number of bona-fide Planetary Mass objects. After applying a completeness correction, we obtain $R$ values in the range 3 to 6, comparable to the star/brown dwarf ratio, and consistent with a steady decrease of the number of cluster members down to planetary masses \citep{2018haex.bookE..92B}.

If we use 0.072~\Msun and 0.012~\Msun as the Hydrogen and Deuterium burning mass limits, both values appropriate for solar metallicity, we count 136 Brown Dwarfs and 34 Planetary Mass Objects with Bayes Factor=2, i.e. bona-fide members of the ONC. 
These numbers combined are comparable to the total number of known low-gravity L-dwarfs  \citep{2018haex.bookE..92B, 2018haex.bookE.118B}, making our catalog one of the richest census to-date of substellar and planetary mass objects in a nearby star forming region.

\begin{deluxetable*}{cccccc}
\tabletypesize{\scriptsize}
\tablewidth{0pt}




\tablecaption{Radial dependence of the number of ONC sources
\label{tbl:Final Counts}}

\tablehead{\colhead{Radial zone} & \colhead{Areal coverage } & \colhead{selection}& \colhead{Stars} & \colhead{Brown Dwarfs} & \colhead{Planetary Mass Objects} \\
                               &  &  (sq.arcsec)             
                               & $ 0.08\le M<1.00~\Msun$ 
                               & $ 0.015\le M<0.03~\Msun$  
                               & $ 0.005\le M<0.015~\Msun$ 
} 

\startdata
$d\leq0.35pc$   & 328   & Total             & 430 & 110 & 245 \\
                &       & Bayes Factor =2   & 300 & 46 & 25        \\
$0.35<\leq0.7pc$&  128  & Total             & 311 & 85 & 405 \\
                &       & Bayes Factor =2   & 216 & 34 & 9\\
$0.7<\leq1.1pc$&  1283  & Total              & 268 & 73 & 703 \\
                &       & Bayes Factor =2   & 205 & 27 & 9\\
$0.7<\leq1.1pc$ & 1983  & Total             & 264 & 61 & 1453 \\
                &       & Bayes Factor =2   & 188 & 26 & 19\\
\enddata
\end{deluxetable*}

\begin{figure}[htb!]
\begin{center}
\includegraphics[width=90mm,angle=0]{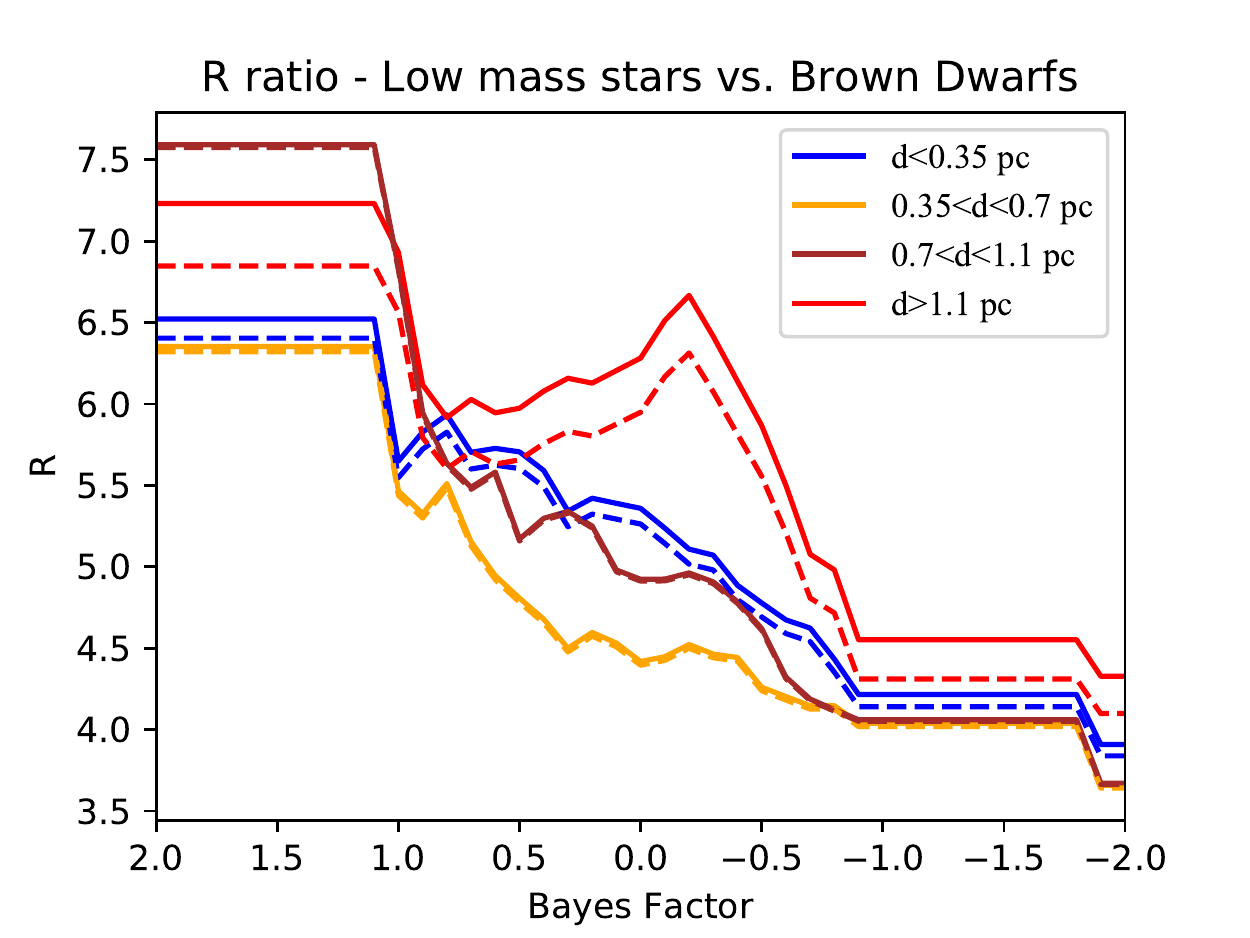}
\caption{\textit{Star (0.08-1.0~\Msun) to brown dwarf (0.03-0.08~\Msun) ratio $R$ vs. Bayes Factor for the four radial zones indicated in the label. Solid lines refer to the observed counts, dashed lines refer to the counts after completeness correction.} \label{Fig:Rvsdistance}}
\end{center}
\end{figure}

\begin{figure}[htb!]
\begin{center}
\includegraphics[width=90mm,angle=0]{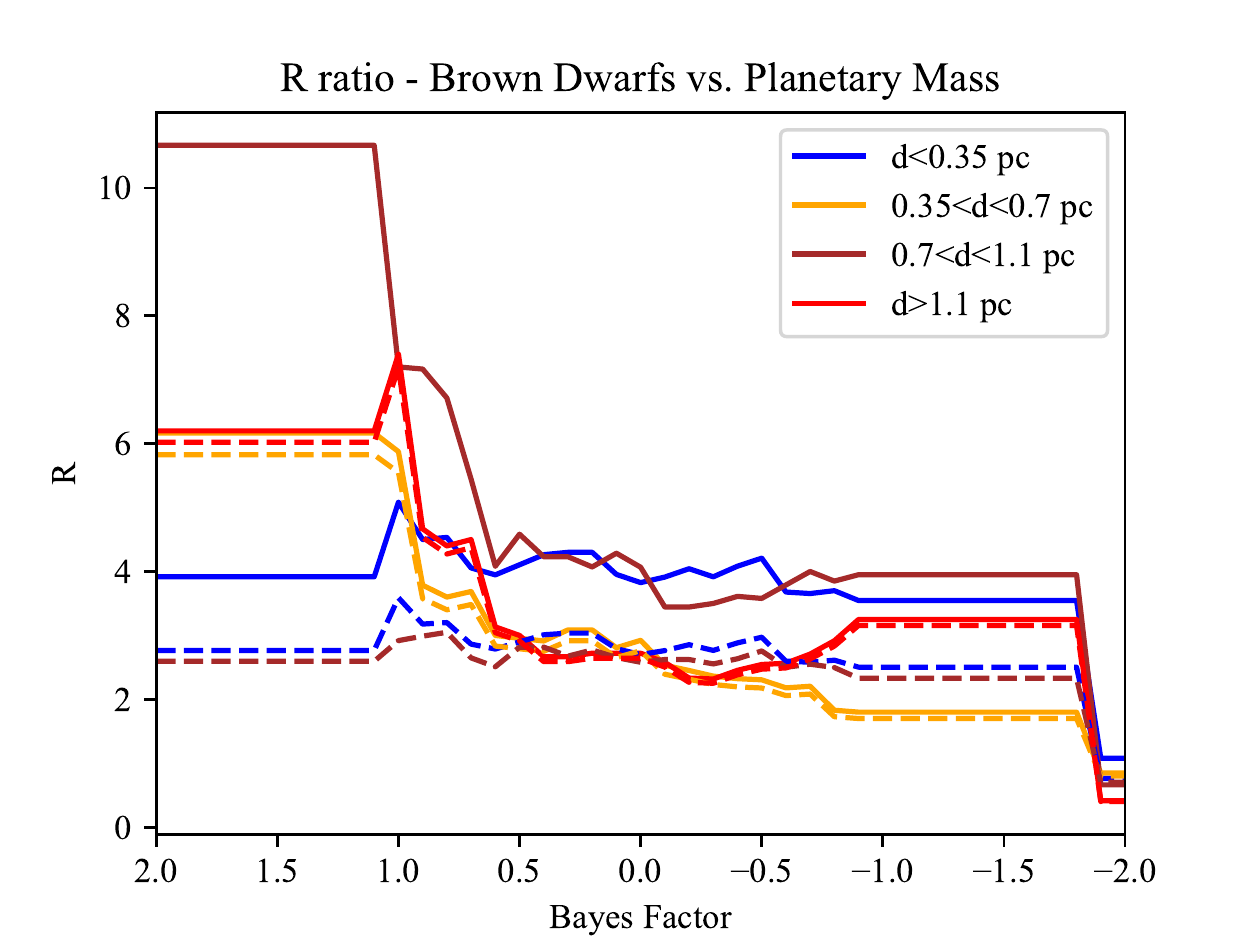}
\caption{\textit{Same as Figure~\ref{Fig:Rvsdistance} for the Brown dwarf (0.03-0.08~\Msun) to planetary mass objects (0.005-0.03~\Msun).} \label{Fig:Rvsdistance_Planets}}
\end{center}
\end{figure}

}

\subsection{Luminosity and Initial Mass Function}
{

In Figure~\ref{Fig:IMF} we present the $m_{130}$ luminosity function, and the corresponding initial mass function, for our four radial regions centered on $\theta^1$Ori-C, limited to our bona-fide cluster members (Bayes Factor = 2) and including completeness correction. The distribution, for all regions, peaks at $m_{130}\sim 12-13$, corresponding to about 0.4-0.2~\Msun. The rapid decrease at $m_{130}<12$ is due to our saturation limit. However, the decrease af fainter magnitudes, $m_{130}>14$, in the substellar regime, is real. 

In general, the discrepancy with e.g. \cite{2011A&A...534A..10A}, i.e. our lower $R$ ratio, can be reconciled if one accepts sources with low Bayes Factor. In other words, if one increases the fraction of contaminants. These difference can thus be be regarded as just another proof of the difficulty of discriminating the true population of substellar and planetary mass objects in the ONC from the reddened population of galactic and extragalactic contaminants using  broad-band photometry plus correction for field star contamination. We defer to Paper~II the detailed analysis of the low-mass IMF of the ONC.

\begin{figure*}[htb!]
\begin{center}
\includegraphics[width=90mm,angle=0]{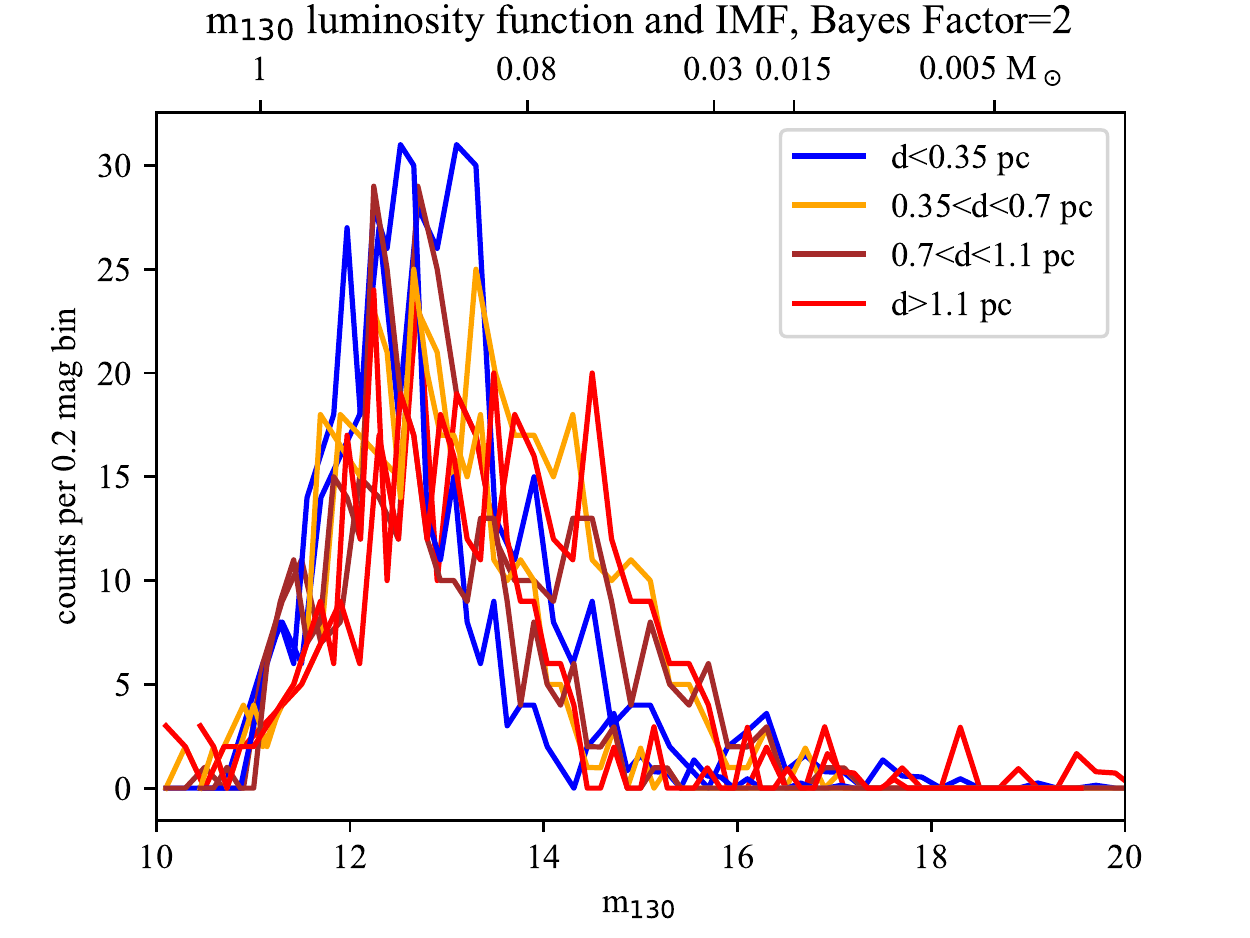}
\caption{\textit{$m_{130}$ luminosity function for our four radial regions centered on $\theta^1 Ori-C$, limited to our bona-fide cluster members (Bayes Factor = 2) and including completeness correction. The top axis indicates the corresponding stellar masses according to our 1~Myr isochrone.} \label{Fig:IMF}}
\end{center}
\end{figure*}

} 

\begin{deluxetable}{ccccc}
\tabletypesize{\scriptsize}
\tablecaption{Estimated parameters of the ONC sample\label{Tab:Calalog_AMT_BR}}
\tablehead{\colhead{Index} 
  & \colhead{$\log$(Bayes Factor)} 
     & \colhead{Mass}
         & \colhead{\teff}
            & \colhead{$A_V$}
                \\ 
  &  & \colhead($M_\odot$) & \colhead{(K)} & \colhead{(mag)}} 
\startdata
   1  & 2.00 & 0.91 & 3757 &  -3.62 \\
   3  & 2.00 & 0.81 & 3678 &   3.30 \\
   4  & 2.00 & 0.28 & 3266 &   0.12 \\
   5  & 2.00 & 0.14 & 3084 &   1.7 \\ 
...	 & ...	 & ...	     & ...	 & ...\\  
\enddata
\tablecomments{Table~\ref{Tab:Calalog_AMT_BR} 
is published in its entirety in the machine-readable format. A portion is shown here for guidance regarding its form and content.}
\end{deluxetable}

\section{Summary}
We have carried out a Hubble Treasury Program (GO-13826) to determine a complete and unbiased census of the stellar population of the Orion Nebula Cluster down to a few Jupiter masses. To disentangle young sub-stellar sources from the background galactic population and extragalactic sources, we have searched for the presence of H$_2$O vapor in absorption, a key signature of the atmosphere of sub-stellar objects. Using the  infrared channel of the Wide-Field-Camera 3 of the Hubble Space Telescope in the 
F139M and F130N  filters,  respectively corresponding to  the  $1.4~\mu$m H$_2$O  absorption feature  and adjacent line-free continuum, we have built a spectral index sensitive to the presence and depth of atmospheric water vapour. We have detected 4504 sources down to magnitudes $\simeq 21.5-22$, corresponding to masses  $\sim3$M$_{Jup}$ for a 1~Myr isochrone at $\simeq 400$~pc. 
About 742  sources, about 1/6th of the total, are fainter that $m_{130}$=14 and have a negative F130M-F139N color index, indicative of the presence of H$_2$O vapor in absorption. These sources can be classified as  bona-fide { M and L} dwarfs, with  temperatures  $T\lesssim 2,850$~K at an assumed  1~Myr cluster age. Their locus on the
color-magnitude diagram is clearly distinct from the larger background population of highly reddened stars and galaxies with positive F130M-F139N color index. The faintest sources with H$_2$O  absorption can be traced down to the sensitivity limit of our survey. We present the source photometry, an estimate of the main stellar parameters and the membership probability based on a Bayesian analysis of our sample. The implications of our measures on the low-mass tail of the Initial Mass Function, determined taking account all ancillary information on the cluster structure, is presented in Paper-II of this series.

\acknowledgments
Support for Program number GO-13826 was provided by NASA through a grant from the Space Telescope Science Institute, which is operated by the Association of Universities for Research in Astronomy, Incorporated, under NASA contract NASS-26555. CFM acknowledges an ESO fellowship. The authors are indebted to an anonymous referee for useful comments and wish to thank Mike Meyer for reviewing an earlier version of the manuscript. This research has made use of the VizieR catalogue access tool, CDS,  Strasbourg, France. The original description of the VizieR service was published in A\&AS 143, 23.

\bibliography{references.bib}

\begin{thebibliography}{}
\expandafter\ifx\csname natexlab\endcsname\relax\def\natexlab#1{#1}\fi
\providecommand{\url}[1]{\href{#1}{#1}}
\providecommand{\dodoi}[1]{doi:~\href{http://doi.org/#1}{\nolinkurl{#1}}}
\providecommand{\doeprint}[1]{\href{http://ascl.net/#1}{\nolinkurl{http://ascl.net/#1}}}
\providecommand{\doarXiv}[1]{\href{https://arxiv.org/abs/#1}{\nolinkurl{https://arxiv.org/abs/#1}}}

\bibitem[{{Abel} {et~al.}(2019){Abel}, {Ferland}, \&
  {O{\textquoteright}Dell}}]{2019ApJ...881..130A}
{Abel}, N.~P., {Ferland}, G.~J., \& {O{\textquoteright}Dell}, C.~R. 2019, \apj,
  881, 130, \dodoi{10.3847/1538-4357/ab2a6e}

\bibitem[{{Ackerman} \& {Marley}(2001)}]{2001ApJ...556..872A}
{Ackerman}, A.~S., \& {Marley}, M.~S. 2001, \apj, 556, 872,
  \dodoi{10.1086/321540}

\bibitem[{{Andersen} {et~al.}(2008){Andersen}, {Meyer}, {Greissl}, \&
  {Aversa}}]{2008ApJ...683L.183A}
{Andersen}, M., {Meyer}, M.~R., {Greissl}, J., \& {Aversa}, A. 2008, \apjl,
  683, L183, \dodoi{10.1086/591473}

\bibitem[{{Andersen} {et~al.}(2011){Andersen}, {Meyer}, {Robberto}, {Bergeron},
  \& {Reid}}]{2011A&A...534A..10A}
{Andersen}, M., {Meyer}, M.~R., {Robberto}, M., {Bergeron}, L.~E., \& {Reid},
  N. 2011, \aap, 534, A10, \dodoi{10.1051/0004-6361/201117062}

\bibitem[{{Bally} {et~al.}(2000){Bally}, {O'Dell}, \&
  {McCaughrean}}]{2000AJ....119.2919B}
{Bally}, J., {O'Dell}, C.~R., \& {McCaughrean}, M.~J. 2000, \aj, 119, 2919,
  \dodoi{10.1086/301385}

\bibitem[{{B{\'e}jar} \& {Mart{\'\i}n}(2018)}]{2018haex.bookE..92B}
{B{\'e}jar}, V.~J.~S., \& {Mart{\'\i}n}, E.~L. 2018, in Handbook of Exoplanets,
  92, \dodoi{10.1007/978-3-319-55333-7_92}

\bibitem[{{Bouy} {et~al.}(2014){Bouy}, {Alves}, {Bertin}, {Sarro}, \&
  {Barrado}}]{2014A&A...564A..29B}
{Bouy}, H., {Alves}, J., {Bertin}, E., {Sarro}, L.~M., \& {Barrado}, D. 2014,
  \aap, 564, A29, \dodoi{10.1051/0004-6361/201323191}

\bibitem[{{Burningham}(2018)}]{2018haex.bookE.118B}
{Burningham}, B. 2018, in Handbook of Exoplanets, 118,
  \dodoi{10.1007/978-3-319-55333-7_118}

\bibitem[{{Burrows} {et~al.}(2003){Burrows}, {Sudarsky}, \&
  {Lunine}}]{2003ApJ...596..587B}
{Burrows}, A., {Sudarsky}, D., \& {Lunine}, J.~I. 2003, \apj, 596, 587,
  \dodoi{10.1086/377709}

\bibitem[{{Cardelli} {et~al.}(1989){Cardelli}, {Clayton}, \&
  {Mathis}}]{1989ApJ...345..245C}
{Cardelli}, J.~A., {Clayton}, G.~C., \& {Mathis}, J.~S. 1989, \apj, 345, 245,
  \dodoi{10.1086/167900}

\bibitem[{{Chabrier} \& {Baraffe}(2000)}]{2000ARA&A..38..337C}
{Chabrier}, G., \& {Baraffe}, I. 2000, \araa, 38, 337,
  \dodoi{10.1146/annurev.astro.38.1.337}

\bibitem[{{Cook} {et~al.}(2017){Cook}, {Scholz}, \&
  {Jayawardhana}}]{2017AJ....154..256C}
{Cook}, N.~J., {Scholz}, A., \& {Jayawardhana}, R. 2017, \aj, 154, 256,
  \dodoi{10.3847/1538-3881/aa9751}

\bibitem[{{Correnti} {et~al.}(2016){Correnti}, {Gennaro}, {Kalirai}, {Brown},
  \& {Calamida}}]{2016ApJ...823...18C}
{Correnti}, M., {Gennaro}, M., {Kalirai}, J.~S., {Brown}, T.~M., \& {Calamida},
  A. 2016, \apj, 823, 18, \dodoi{10.3847/0004-637X/823/1/18}

\bibitem[{{Da Rio} {et~al.}(2012){Da Rio}, {Robberto}, {Hillenbrand},
  {Henning}, \& {Stassun}}]{2012ApJ...748...14D}
{Da Rio}, N., {Robberto}, M., {Hillenbrand}, L.~A., {Henning}, T., \&
  {Stassun}, K.~G. 2012, \apj, 748, 14, \dodoi{10.1088/0004-637X/748/1/14}

\bibitem[{{Da Rio} {et~al.}(2010){Da Rio}, {Robberto}, {Soderblom}, {Panagia},
  {Hillenbrand}, {Palla}, \& {Stassun}}]{2010ApJ...722.1092D}
{Da Rio}, N., {Robberto}, M., {Soderblom}, D.~R., {et~al.} 2010, \apj, 722,
  1092, \dodoi{10.1088/0004-637X/722/2/1092}

\bibitem[{{Da Rio} {et~al.}(2016){Da Rio}, {Tan}, {Covey}, {Cottaar}, {Foster},
  {Cullen}, {Tobin}, {Kim}, {Meyer}, {Nidever}, {Stassun}, {Chojnowski},
  {Flaherty}, {Majewski}, {Skrutskie}, {Zasowski}, \&
  {Pan}}]{2016ApJ...818...59D}
{Da Rio}, N., {Tan}, J.~C., {Covey}, K.~R., {et~al.} 2016, \apj, 818, 59,
  \dodoi{10.3847/0004-637X/818/1/59}

\bibitem[{{Dalcanton} {et~al.}(2012){Dalcanton}, {Williams}, {Lang}, {Lauer},
  {Kalirai}, {Seth}, {Dolphin}, {Rosenfield}, {Weisz}, {Bell}, {Bianchi},
  {Boyer}, {Caldwell}, {Dong}, {Dorman}, {Gilbert}, {Girardi}, {Gogarten},
  {Gordon}, {Guhathakurta}, {Hodge}, {Holtzman}, {Johnson}, {Larsen}, {Lewis},
  {Melbourne}, {Olsen}, {Rix}, {Rosema}, {Saha}, {Sarajedini}, {Skillman}, \&
  {Stanek}}]{2012ApJS..200...18D}
{Dalcanton}, J.~J., {Williams}, B.~F., {Lang}, D., {et~al.} 2012, \apjs, 200,
  18, \dodoi{10.1088/0067-0049/200/2/18}

\bibitem[{{Delfosse} {et~al.}(1999){Delfosse}, {Tinney}, {Forveille},
  {Epchtein}, {Borsenberger}, {Fouqu{\'e}}, {Kimeswenger}, \&
  {Tiph{\`e}ne}}]{1999A&AS..135...41D}
{Delfosse}, X., {Tinney}, C.~G., {Forveille}, T., {et~al.} 1999, \aaps, 135,
  41, \dodoi{10.1051/aas:1999158}

\bibitem[{{Dolphin}(2000)}]{2000PASP..112.1383D}
{Dolphin}, A.~E. 2000, \pasp, 112, 1383, \dodoi{10.1086/316630}

\bibitem[{{Drass} {et~al.}(2016){Drass}, {Haas}, {Chini}, {Bayo}, {Hackstein},
  {Hoffmeister}, {Godoy}, \& {Vogt}}]{2016MNRAS.461.1734D}
{Drass}, H., {Haas}, M., {Chini}, R., {et~al.} 2016, \mnras, 461, 1734,
  \dodoi{10.1093/mnras/stw1094}

\bibitem[{{Dressel}(2014)}]{Dressel2014}
{Dressel}, L. 2014, {Wide Field Camera 3 Instrument Handbook, Version 6.0
  (Baltimore: STScI)}, Tech. rep.

\bibitem[{{Eisner} {et~al.}(2016){Eisner}, {Bally}, {Ginsburg}, \&
  {Sheehan}}]{2016ApJ...826...16E}
{Eisner}, J.~A., {Bally}, J.~M., {Ginsburg}, A., \& {Sheehan}, P.~D. 2016,
  \apj, 826, 16, \dodoi{10.3847/0004-637X/826/1/16}

\bibitem[{{Faherty} {et~al.}(2016){Faherty}, {Riedel}, {Cruz}, {Gagne},
  {Filippazzo}, {Lambrides}, {Fica}, {Weinberger}, {Thorstensen}, {Tinney},
  {Baldassare}, {Lemonier}, \& {Rice}}]{2016ApJS..225...10F}
{Faherty}, J.~K., {Riedel}, A.~R., {Cruz}, K.~L., {et~al.} 2016, \apjs, 225,
  10, \dodoi{10.3847/0067-0049/225/1/10}

\bibitem[{{Gorlova} {et~al.}(2003){Gorlova}, {Meyer}, {Rieke}, \&
  {Liebert}}]{2003ApJ...593.1074G}
{Gorlova}, N.~I., {Meyer}, M.~R., {Rieke}, G.~H., \& {Liebert}, J. 2003, \apj,
  593, 1074, \dodoi{10.1086/376730}

\bibitem[{{Gro\ss{}schedl} {et~al.}(2018){Gro\ss{}schedl}, {Alves}, {Meingast},
  {Ackerl}, {Ascenso}, {Bouy}, {Burkert}, {Forbrich}, {F\"urnkranz}, {Goodman},
  {Hacar}, {Herbst-Kiss}, {Lada, C., and {Larreina}, I., and {Leschinski}, K.,
  and {Lombardi}, M., and {Moitinho}, A.,}, {Mortimer}, \& {Zari}}]{refId0}
{Gro\ss{}schedl}, J.~E., {Alves}, J., {Meingast}, S., {et~al.} 2018, A\&A, 619,
  A106, \dodoi{10.1051/0004-6361/201833901}

\bibitem[{{Herbst} {et~al.}(2002){Herbst}, {Bailer-Jones}, {Mundt},
  {Meisenheimer}, \& {Wackermann}}]{2002A&A...396..513H}
{Herbst}, W., {Bailer-Jones}, C.~A.~L., {Mundt}, R., {Meisenheimer}, K., \&
  {Wackermann}, R. 2002, \aap, 396, 513, \dodoi{10.1051/0004-6361:20021362}

\bibitem[{{Hillenbrand} \& {Carpenter}(2000)}]{2000ApJ...540..236H}
{Hillenbrand}, L.~A., \& {Carpenter}, J.~M. 2000, \apj, 540, 236,
  \dodoi{10.1086/309309}

\bibitem[{{Hillenbrand} {et~al.}(2013){Hillenbrand}, {Hoffer}, \&
  {Herczeg}}]{2013AJ....146...85H}
{Hillenbrand}, L.~A., {Hoffer}, A.~S., \& {Herczeg}, G.~J. 2013, \aj, 146, 85,
  \dodoi{10.1088/0004-6256/146/4/85}

\bibitem[{{Husser} {et~al.}(2013){Husser}, {Wende-von Berg}, {Dreizler},
  {Homeier}, {Reiners}, {Barman}, \& {Hauschildt}}]{2013A&A...553A...6H}
{Husser}, T.~O., {Wende-von Berg}, S., {Dreizler}, S., {et~al.} 2013, \aap,
  553, A6, \dodoi{10.1051/0004-6361/201219058}

\bibitem[{{Ingraham} {et~al.}(2014){Ingraham}, {Albert}, {Doyon}, \&
  {Artigau}}]{2014ApJ...782....8I}
{Ingraham}, P., {Albert}, L., {Doyon}, R., \& {Artigau}, E. 2014, \apj, 782, 8,
  \dodoi{10.1088/0004-637X/782/1/8}

\bibitem[{{Jeffries} {et~al.}(2011){Jeffries}, {Littlefair}, {Naylor}, \&
  {Mayne}}]{2011MNRAS.418.1948J}
{Jeffries}, R.~D., {Littlefair}, S.~P., {Naylor}, T., \& {Mayne}, N.~J. 2011,
  \mnras, 418, 1948, \dodoi{10.1111/j.1365-2966.2011.19613.x}

\bibitem[{{Jones} {et~al.}(1994){Jones}, {Longmore}, {Jameson}, \&
  {Mountain}}]{1994MNRAS.267..413J}
{Jones}, H.~R.~A., {Longmore}, A.~J., {Jameson}, R.~F., \& {Mountain}, C.~M.
  1994, \mnras, 267, 413, \dodoi{10.1093/mnras/267.2.413}

\bibitem[{Kirkpatrick {et~al.}(1999)Kirkpatrick, Reid, Liebert, Cutri, Nelson,
  Beichman, Dahn, Monet, Gizis, \& Skrutskie}]{Kirkpatrick_1999}
Kirkpatrick, J.~D., Reid, I.~N., Liebert, J., {et~al.} 1999, The Astrophysical
  Journal, 519, 802, \dodoi{10.1086/307414}

\bibitem[{{Koekemoer} {et~al.}(2011){Koekemoer}, {Faber}, {Ferguson}, {Grogin},
  {Kocevski}, {Koo}, {Lai}, {Lotz}, {Lucas}, {McGrath}, {Ogaz}, {Rajan},
  {Riess}, {Rodney}, {Strolger}, {Casertano}, {Castellano}, {Dahlen},
  {Dickinson}, {Dolch}, {Fontana}, {Giavalisco}, {Grazian}, {Guo}, {Hathi},
  {Huang}, {van der Wel}, {Yan}, {Acquaviva}, {Alexander}, {Almaini}, {Ashby},
  {Barden}, {Bell}, {Bournaud}, {Brown}, {Caputi}, {Cassata}, {Challis},
  {Chary}, {Cheung}, {Cirasuolo}, {Conselice}, {Roshan Cooray}, {Croton},
  {Daddi}, {Dav{\'e}}, {de Mello}, {de Ravel}, {Dekel}, {Donley}, {Dunlop},
  {Dutton}, {Elbaz}, {Fazio}, {Filippenko}, {Finkelstein}, {Frazer}, {Gardner},
  {Garnavich}, {Gawiser}, {Gruetzbauch}, {Hartley}, {H{\"a}ussler},
  {Herrington}, {Hopkins}, {Huang}, {Jha}, {Johnson}, {Kartaltepe},
  {Khostovan}, {Kirshner}, {Lani}, {Lee}, {Li}, {Madau}, {McCarthy},
  {McIntosh}, {McLure}, {McPartland}, {Mobasher}, {Moreira}, {Mortlock},
  {Moustakas}, {Mozena}, {Nandra}, {Newman}, {Nielsen}, {Niemi}, {Noeske},
  {Papovich}, {Pentericci}, {Pope}, {Primack}, {Ravindranath}, {Reddy},
  {Renzini}, {Rix}, {Robaina}, {Rosario}, {Rosati}, {Salimbeni}, {Scarlata},
  {Siana}, {Simard}, {Smidt}, {Snyder}, {Somerville}, {Spinrad}, {Straughn},
  {Telford}, {Teplitz}, {Trump}, {Vargas}, {Villforth}, {Wagner}, {Wandro},
  {Wechsler}, {Weiner}, {Wiklind}, {Wild}, {Wilson}, {Wuyts}, \&
  {Yun}}]{2011ApJS..197...36K}
{Koekemoer}, A.~M., {Faber}, S.~M., {Ferguson}, H.~C., {et~al.} 2011, \apjs,
  197, 36, \dodoi{10.1088/0067-0049/197/2/36}

\bibitem[{{Kong} {et~al.}(2018){Kong}, {Arce}, {Feddersen}, {Carpenter},
  {Nakamura}, {Shimajiri}, {Isella}, {Ossenkopf-Okada}, {Sargent},
  {S{\'a}nchez-Monge}, {Suri}, {Kauffmann}, {Pillai}, {Pineda}, {Koda},
  {Bally}, {Lis}, {Padoan}, {Klessen}, {Mairs}, {Goodman}, {Goldsmith},
  {McGehee}, {Schilke}, {Teuben}, {Maureira}, {Hara}, {Ginsburg}, {Burkhart},
  {Smith}, {Schmiedeke}, {Pineda}, {Ishii}, {Sasaki}, {Kawabe}, {Urasawa},
  {Oyamada}, \& {Tanabe}}]{2018ApJS..236...25K}
{Kong}, S., {Arce}, H.~G., {Feddersen}, J.~R., {et~al.} 2018, \apjs, 236, 25,
  \dodoi{10.3847/1538-4365/aabafc}

\bibitem[{{Kounkel} {et~al.}(2017){Kounkel}, {Hartmann}, {Loinard},
  {Ortiz-Le{\'o}n}, {Mioduszewski}, {Rodr{\'{\i}}guez}, {Dzib}, {Torres},
  {Pech}, {Galli}, {Rivera}, {Boden}, {Evans}, {Brice{\~n}o}, \&
  {Tobin}}]{2017ApJ...834..142K}
{Kounkel}, M., {Hartmann}, L., {Loinard}, L., {et~al.} 2017, \apj, 834, 142,
  \dodoi{10.3847/1538-4357/834/2/142}

\bibitem[{{Kounkel} {et~al.}(2018){Kounkel}, {Covey}, {Su{\'a}rez},
  {Rom{\'a}n-Z{\'u}{\~n}iga}, {Hernandez}, {Stassun}, {Jaehnig}, {Feigelson},
  {Pe{\~n}a Ram{\'\i}rez}, {Roman-Lopes}, {Da Rio}, {Stringfellow}, {Kim},
  {Borissova}, {Fern{\'a}ndez- Trincado}, {Burgasser},
  {Garc{\'\i}a-Hern{\'a}ndez}, {Zamora}, {Pan}, \&
  {Nitschelm}}]{2018AJ....156...84K}
{Kounkel}, M., {Covey}, K., {Su{\'a}rez}, G., {et~al.} 2018, \aj, 156, 84,
  \dodoi{10.3847/1538-3881/aad1f1}

\bibitem[{{Krist} {et~al.}(2011){Krist}, {Hook}, \&
  {Stoehr}}]{2011SPIE.8127E..0JK}
{Krist}, J.~E., {Hook}, R.~N., \& {Stoehr}, F. 2011, in \procspie, Vol. 8127,
  Optical Modeling and Performance Predictions V, 81270J,
  \dodoi{10.1117/12.892762}

\bibitem[{{Kritsuk} {et~al.}(2007){Kritsuk}, {Norman}, {Padoan}, \&
  {Wagner}}]{2007ApJ...665..416K}
{Kritsuk}, A.~G., {Norman}, M.~L., {Padoan}, P., \& {Wagner}, R. 2007, \apj,
  665, 416, \dodoi{10.1086/519443}

\bibitem[{{Kritsuk} {et~al.}(2017){Kritsuk}, {Ustyugov}, \&
  {Norman}}]{2017NJPh...19f5003K}
{Kritsuk}, A.~G., {Ustyugov}, S.~D., \& {Norman}, M.~L. 2017, New Journal of
  Physics, 19, 065003, \dodoi{10.1088/1367-2630/aa7156}

\bibitem[{{Kroupa}(2001)}]{2001MNRAS.322..231K}
{Kroupa}, P. 2001, \mnras, 322, 231, \dodoi{10.1046/j.1365-8711.2001.04022.x}

\bibitem[{{Kuhn} {et~al.}(2019){Kuhn}, {Hillenbrand}, {Sills}, {Feigelson}, \&
  {Getman}}]{2019ApJ...870...32K}
{Kuhn}, M.~A., {Hillenbrand}, L.~A., {Sills}, A., {Feigelson}, E.~D., \&
  {Getman}, K.~V. 2019, \apj, 870, 32, \dodoi{10.3847/1538-4357/aaef8c}

\bibitem[{{Lada} \& {Lada}(2003)}]{2003ARA&A..41...57L}
{Lada}, C.~J., \& {Lada}, E.~A. 2003, \araa, 41, 57,
  \dodoi{10.1146/annurev.astro.41.011802.094844}

\bibitem[{{Laidler}(2008)}]{Laidler+08}
{Laidler}, V. e.~a. 2008, {Synphot Data User's Guide (Baltimore: STScI)}, Tech.
  rep.

\bibitem[{{Lucas} {et~al.}(2005){Lucas}, {Roche}, \&
  {Tamura}}]{2005MNRAS.361..211L}
{Lucas}, P.~W., {Roche}, P.~F., \& {Tamura}, M. 2005, \mnras, 361, 211,
  \dodoi{10.1111/j.1365-2966.2005.09156.x}

\bibitem[{{Luhman} {et~al.}(2003){Luhman}, {Stauffer}, {Muench}, {Rieke},
  {Lada}, {Bouvier}, \& {Lada}}]{2003ApJ...593.1093L}
{Luhman}, K.~L., {Stauffer}, J.~R., {Muench}, A.~A., {et~al.} 2003, \apj, 593,
  1093, \dodoi{10.1086/376594}

\bibitem[{{Manara} {et~al.}(2012){Manara}, {Robberto}, {Da Rio}, {Lodato},
  {Hillenbrand}, {Stassun}, \& {Soderblom}}]{2012ApJ...755..154M}
{Manara}, C.~F., {Robberto}, M., {Da Rio}, N., {et~al.} 2012, \apj, 755, 154,
  \dodoi{10.1088/0004-637X/755/2/154}

\bibitem[{{Menten} {et~al.}(2007){Menten}, {Reid}, {Forbrich}, \&
  {Brunthaler}}]{2007A&A...474..515M}
{Menten}, K.~M., {Reid}, M.~J., {Forbrich}, J., \& {Brunthaler}, A. 2007, \aap,
  474, 515, \dodoi{10.1051/0004-6361:20078247}

\bibitem[{{Mohanty} {et~al.}(2005){Mohanty}, {Basri}, \&
  {Jayawardhana}}]{2005AN....326..891M}
{Mohanty}, S., {Basri}, G., \& {Jayawardhana}, R. 2005, Astronomische
  Nachrichten, 326, 891, \dodoi{10.1002/asna.200510450}

\bibitem[{{Morales-Calder{\'o}n} {et~al.}(2012){Morales-Calder{\'o}n},
  {Stauffer}, {Stassun}, {Vrba}, {Prato}, {Hillenbrand}, {Terebey}, {Covey},
  {Rebull}, {Terndrup}, {Gutermuth}, {Song}, {Plavchan}, {Carpenter},
  {Marchis}, {Garc{\'\i}a}, {Margheim}, {Luhman}, {Angione}, \&
  {Irwin}}]{2012ApJ...753..149M}
{Morales-Calder{\'o}n}, M., {Stauffer}, J.~R., {Stassun}, K.~G., {et~al.} 2012,
  \apj, 753, 149, \dodoi{10.1088/0004-637X/753/2/149}

\bibitem[{{Morgan} {et~al.}(1943){Morgan}, {Keenan}, \&
  {Kellman}}]{1943assw.book.....M}
{Morgan}, W.~W., {Keenan}, P.~C., \& {Kellman}, E. 1943, {An atlas of stellar
  spectra, with an outline of spectral classification}

\bibitem[{{Morley} {et~al.}(2014){Morley}, {Marley}, {Fortney}, {Lupu},
  {Saumon}, {Greene}, \& {Lodders}}]{2014ApJ...787...78M}
{Morley}, C.~V., {Marley}, M.~S., {Fortney}, J.~J., {et~al.} 2014, \apj, 787,
  78, \dodoi{10.1088/0004-637X/787/1/78}

\bibitem[{{Muench} {et~al.}(2008){Muench}, {Getman}, {Hillenbrand}, \&
  {Preibisch}}]{2008hsf1.book..483M}
{Muench}, A., {Getman}, K., {Hillenbrand}, L., \& {Preibisch}, T. 2008, {Star
  Formation in the Orion Nebula I: Stellar Content}, ed. B.~{Reipurth}, 483

\bibitem[{{Muench} {et~al.}(2002){Muench}, {Lada}, {Lada}, \&
  {Alves}}]{2002ApJ...573..366M}
{Muench}, A.~A., {Lada}, E.~A., {Lada}, C.~J., \& {Alves}, J. 2002, \apj, 573,
  366, \dodoi{10.1086/340554}

\bibitem[{{Mu{\v{z}}i{\'c}} {et~al.}(2019){Mu{\v{z}}i{\'c}}, {Scholz},
  {Pe{\~n}a Ram{\'\i}rez}, {Jayawardhana}, {Sch{\"o}del}, {Geers}, {Cieza}, \&
  {Bayo}}]{2019ApJ...881...79}
{Mu{\v{z}}i{\'c}}, K., {Scholz}, A., {Pe{\~n}a Ram{\'\i}rez}, K., {et~al.}
  2019, \apj, 881, 79, \dodoi{10.3847/1538-4357/ab2da4}

\bibitem[{{Najita} {et~al.}(2000){Najita}, {Tiede}, \&
  {Carr}}]{2000ApJ...541..977N}
{Najita}, J.~R., {Tiede}, G.~P., \& {Carr}, J.~S. 2000, \apj, 541, 977,
  \dodoi{10.1086/309477}

\bibitem[{{O'Dell} {et~al.}(2008){O'Dell}, {Muench}, {Smith}, \&
  {Zapata}}]{2008hsf1.book..544O}
{O'Dell}, C.~R., {Muench}, A., {Smith}, N., \& {Zapata}, L. 2008, {Star
  Formation in the Orion Nebula II: Gas, Dust, Proplyds and Outflows}, ed.
  B.~{Reipurth}, 544

\bibitem[{{O'dell} \& {Wong}(1996)}]{1996AJ....111..846O}
{O'dell}, C.~R., \& {Wong}, K. 1996, \aj, 111, 846, \dodoi{10.1086/117832}

\bibitem[{{Pacifici} {et~al.}(2016){Pacifici}, {Kassin}, {Weiner}, {Holden},
  {Gardner}, {Faber}, {Ferguson}, {Koo}, {Primack}, {Bell}, {Dekel}, {Gawiser},
  {Giavalisco}, {Rafelski}, {Simons}, {Barro}, {Croton}, {Dav{\'e}}, {Fontana},
  {Grogin}, {Koekemoer}, {Lee}, {Salmon}, {Somerville}, \&
  {Behroozi}}]{2016ApJ...832...79P}
{Pacifici}, C., {Kassin}, S.~A., {Weiner}, B.~J., {et~al.} 2016, \apj, 832, 79,
  \dodoi{10.3847/0004-637X/832/1/79}

\bibitem[{Portegies~Zwart {et~al.}(2010)Portegies~Zwart, McMillan, \&
  Gieles}]{doi:10.1146/annurev-astro-081309-130834}
Portegies~Zwart, S.~F., McMillan, S.~L., \& Gieles, M. 2010, Annual Review of
  Astronomy and Astrophysics, 48, 431,
  \dodoi{10.1146/annurev-astro-081309-130834}

\bibitem[{{Reggiani} {et~al.}(2011){Reggiani}, {Robberto}, {Da Rio}, {Meyer},
  {Soderblom}, \& {Ricci}}]{2011A&A...534A..83R}
{Reggiani}, M., {Robberto}, M., {Da Rio}, N., {et~al.} 2011, \aap, 534, A83,
  \dodoi{10.1051/0004-6361/201116946}

\bibitem[{{Ricci} {et~al.}(2008){Ricci}, {Robberto}, \&
  {Soderblom}}]{2008AJ....136.2136R}
{Ricci}, L., {Robberto}, M., \& {Soderblom}, D.~R. 2008, \aj, 136, 2136,
  \dodoi{10.1088/0004-6256/136/5/2136}

\bibitem[{{Riddick} {et~al.}(2007){Riddick}, {Roche}, \&
  {Lucas}}]{2007MNRAS.381.1077R}
{Riddick}, F.~C., {Roche}, P.~F., \& {Lucas}, P.~W. 2007, \mnras, 381, 1077,
  \dodoi{10.1111/j.1365-2966.2007.12308.x}

\bibitem[{{Robberto} {et~al.}(2004){Robberto}, {Song}, {Mora Carrillo},
  {Beckwith}, {Makidon}, \& {Panagia}}]{2004ApJ...606..952R}
{Robberto}, M., {Song}, J., {Mora Carrillo}, G., {et~al.} 2004, \apj, 606, 952,
  \dodoi{10.1086/383141}

\bibitem[{{Robberto} {et~al.}(2013){Robberto}, {Soderblom}, {Bergeron},
  {Kozhurina-Platais}, {Makidon}, {McCullough}, {McMaster}, {Panagia}, {Reid},
  {Levay}, {Frattare}, {Da Rio}, {Andersen}, {O'Dell}, {Stassun}, {Simon},
  {Feigelson}, {Stauffer}, {Meyer}, {Reggiani}, {Krist}, {Manara},
  {Romaniello}, {Hillenbrand}, {Ricci}, {Palla}, {Najita}, {Ananna},
  {Scandariato}, \& {Smith}}]{2013ApJS..207...10R}
{Robberto}, M., {Soderblom}, D.~R., {Bergeron}, E., {et~al.} 2013, \apjs, 207,
  10, \dodoi{10.1088/0067-0049/207/1/10}

\bibitem[{{Robin} {et~al.}(2003){Robin}, {Reyl{\'e}}, {Derri{\`e}re}, \&
  {Picaud}}]{2003A&A...409..523R0}
{Robin}, A.~C., {Reyl{\'e}}, C., {Derri{\`e}re}, S., \& {Picaud}, S. 2003,
  \aap, 409, 523, \dodoi{10.1051/0004-6361:20031117}

\bibitem[{{Ruffio} {et~al.}(2018){Ruffio}, {Mawet}, {Czekala}, {Macintosh}, {De
  Rosa}, {Ruane}, {Bottom}, {Pueyo}, {Wang}, {Hirsch}, {Zhu}, \&
  {Nielsen}}]{2018AJ....156..196R}
{Ruffio}, J.-B., {Mawet}, D., {Czekala}, I., {et~al.} 2018, \aj, 156, 196,
  \dodoi{10.3847/1538-3881/aade95}

\bibitem[{{Scandariato} {et~al.}(2011){Scandariato}, {Robberto}, {Pagano}, \&
  {Hillenbrand}}]{2011AA...533A..38S}
{Scandariato}, G., {Robberto}, M., {Pagano}, I., \& {Hillenbrand}, L.~A. 2011,
  \aap, 533, A38, \dodoi{10.1051/0004-6361/201116554}

\bibitem[{{Scholz} {et~al.}(2012){Scholz}, {Muzic}, {Geers}, {Bonavita},
  {Jayawardhana}, \& {Tamura}}]{2012ApJ...744....6S}
{Scholz}, A., {Muzic}, K., {Geers}, V., {et~al.} 2012, \apj, 744, 6,
  \dodoi{10.1088/0004-637X/744/1/6}

\bibitem[{{Slesnick} {et~al.}(2004){Slesnick}, {Hillenbrand}, \&
  {Carpenter}}]{2004ApJ...610.1045S}
{Slesnick}, C.~L., {Hillenbrand}, L.~A., \& {Carpenter}, J.~M. 2004, \apj, 610,
  1045, \dodoi{10.1086/421898}

\bibitem[{{Smith} {et~al.}(2008){Smith}, {Zavodny}, {Rahmer}, \&
  {Bonati}}]{2008SPIE.7021E..0JS}
{Smith}, R.~M., {Zavodny}, M., {Rahmer}, G., \& {Bonati}, M. 2008, in
  \procspie, Vol. 7021, High Energy, Optical, and Infrared Detectors for
  Astronomy III, 70210J, \dodoi{10.1117/12.789372}

\bibitem[{{Wallace} {et~al.}(2000){Wallace}, {Meyer}, {Hinkle}, \&
  {Edwards}}]{2000ApJ...535..325W}
{Wallace}, L., {Meyer}, M.~R., {Hinkle}, K., \& {Edwards}, S. 2000, \apj, 535,
  325, \dodoi{10.1086/308835}

\bibitem[{{Weights} {et~al.}(2009){Weights}, {Lucas}, {Roche}, {Pinfield}, \&
  {Riddick}}]{2009MNRAS.392..817W}
{Weights}, D.~J., {Lucas}, P.~W., {Roche}, P.~F., {Pinfield}, D.~J., \&
  {Riddick}, F. 2009, \mnras, 392, 817,
  \dodoi{10.1111/j.1365-2966.2008.14096.x}

\end{thebibliography}

\facility{HST (WFC3)}

\appendix
{
\section{BT-Settl Model used in this paper}
The models adopted in this paper have been taken from the grids of BT-Settl model available for download at the web address https://phoenix.ens-lyon.fr/Grids/BT-Settl. They  utilize the stellar interior calculations by Baraffe et al. (1997, 1998, 2003), and the  PHOENIX atmospheric model.  
The BT-Settl models, in particular, implement cloud formation as well as dust sedimentation in the atmospheres of brown dwarfs and planetary-mass objects. Since the library is subject to changes, we present in Table~\ref{tab:BT-Settl} the version we adopted, downloaded in October 2016, limited to the passbands used in this paper. We also add, last two columns, the revised values of the F130N and F139M photometry according to the emprical Ridge-Line, derived as detailed in the following section.

To associate effective temperatures to spectral types, we adopt the parameterization provided by \cite{2003ApJ...593.1093L} for young M-type stars, whereas for L-type we adopt \cite{2016ApJS..225...10F} for young moving groups deriving the values listed in Table~\ref{tab:Faherty}.

}





\begin{deluxetable}{ccccccccccccc}




\tablecaption{BT-Settl.M-0.0-HST-WFC3-1Myr
\label{tab:BT-Settl}}


\tablehead{\colhead{M/M$_\odot$} & \colhead{T$_{eff}$} & \colhead{L/L$_\odot$} & \colhead{log(g)} & \colhead{R} & \colhead{D} & \colhead{Li} & \colhead{m130N} & \colhead{m139M} & \colhead{m850LP$_1$} & \colhead{m850LP$_2$}& \colhead{m130N-RL} & \colhead{m139M-RL}\\ 
\colhead{} & \colhead{(K)} & \colhead{} & \colhead{} & \colhead{($10^9$~cm)} 
& \colhead{} & \colhead{}
& \colhead{} & \colhead{} 
& \colhead{} & \colhead{}
& \colhead{} & \colhead{}} 
\startdata
0.0005 & 628 & -5.37 & 2.65 & 12.25 & 1 & 1 & 17.559 & 23.253 & 21.667 & 21.69  &   19.731	&   20.137\\
0.001 & 942 & -4.72 & 3 & 11.57 & 1 & 1 & 14.562 & 17.701 & 17.513 & 17.524     &   15.556	&   15.968\\
0.002 & 1285 & -4.14 & 3.26 & 12.09 & 1 & 1 & 13.133 & 14.632 & 15.768 & 15.773 &   13.801	&   14.216\\
0.003 & 1553 & -3.75 & 3.37 & 12.99 & 1 & 1 & 12.496 & 13.06 & 15.219 & 15.224  &	13.250	&	13.666\\
0.004 & 1747 & -3.48 & 3.44 & 13.92 & 1 & 1 & 11.263 & 12.216 & 14.237 & 14.242 &	12.265	&	12.681\\
0.005 & 1901 & -3.27 & 3.47 & 14.95 & 1 & 1 & 10.5 & 11.544 & 13.322 & 13.328   &	11.460	&	11.836\\
0.006 & 2004 & -3.13 & 3.5 & 15.88 & 1 & 1 & 9.996 & 11.042 & 12.616 & 12.621   &	10.617	&	11.044\\
0.007 & 2098 & -3 & 3.51 & 16.85 & 1 & 1 & 9.763 & 10.682 & 12.289 & 12.294     &	10.294	&	10.719\\
0.008 & 2159 & -2.89 & 3.52 & 17.96 & 1 & 1 & 9.472 & 10.325 & 11.852 & 11.857  &	9.829	&	10.264\\
0.009 & 2207 & -2.81 & 3.53 & 18.88 & 1 & 1 & 9.25 & 10.05 & 11.522 & 11.527    &	9.482	&	9.924\\
0.01 & 2251 & -2.73 & 3.53 & 19.81 & 1 & 1 & 9.083 & 9.829 & 11.29 & 11.295     &	9.220	&	9.673\\
0.012 & 2321 & -2.61 & 3.54 & 21.4 & 1 & 1 & 8.813 & 9.471 & 10.91 & 10.913     &	8.855	&	9.302\\
0.015 & 2400 & -2.46 & 3.54 & 24.04 & 1 & 1 & 8.436 & 8.993 & 10.39 & 10.393    &	8.466	&	8.865\\
0.02 & 2484 & -2.28 & 3.55 & 27.47 & 1 & 1 & 8.035 & 8.503 & 9.862 & 9.865      &	7.954	&	8.347\\
0.03 & 2598 & -2.17 & 3.69 & 28.38 & 0.98 & 1 & 7.822 & 8.177 & 9.485 & 9.488   &	7.625	&	7.999\\
0.04 & 2746 & -1.94 & 3.68 & 33.27 & 0.735 & 1 & 7.301 & 7.547 & 8.794 & 8.796  &	7.061	&	7.388\\
0.05 & 2768 & -1.64 & 3.49 & 46.38 & 0.99 & 1 & 6.548 & 6.79 & 8.031 & 8.033    &	6.500	&	6.753\\
0.06 & 2824 & -1.6 & 3.57 & 46.27 & 0.605 & 1 & 6.484 & 6.694 & 7.91 & 7.912    &	6.432	&	6.665\\
0.07 & 2853 & -1.48 & 3.53 & 52.41 & 0.56 & 1 & 6.171 & 6.376 & 7.581 & 7.583   &	6.217	&	6.408\\
0.072 & 2858 & -1.45 & 3.52 & 53.68 & 0.555 & 1 & 6.112 & 6.316 & 7.519 & 7.521 &	6.137	&	6.320\\
0.075 & 2833 & -1.39 & 3.46 & 58.93 & 0.915 & 1 & 5.94 & 6.158 & 7.375 & 7.377  &	6.002	&	6.179\\
0.08 & 2869 & -1.37 & 3.49 & 58.47 & 0.51 & 1 & 5.907 & 6.111 & 7.311 & 7.313   &	5.939	&	6.115\\
0.09 & 2867 & -1.29 & 3.46 & 64.54 & 0.4895 & 1 & 5.693 & 5.902 & 7.105 & 7.107 &	5.730	&	5.908\\
0.1 & 2856 & -1.19 & 3.39 & 73.18 & 0.8 & 1 & 5.432 & 5.652 & 6.865 & 6.867     &	5.501	&	5.671\\
0.11 & 3023 & -1.08 & 3.43 & 73.48 & 0.312 & 1 & 5.25 & 5.371 & 6.542 & 6.544   &	5.222	&	5.361\\
0.13 & 3072 & -0.96 & 3.41 & 82.37 & 0.1975 & 1 & 4.948 & 5.046 & 6.222 & 6.223 &	4.955	&	5.055\\
0.15 & 3120 & -0.85 & 3.39 & 90.05 & 0.112 & 1 & 4.708 & 4.779 & 5.957 & 5.958  &	4.732	&	4.801\\
0.175 & 3163 & -0.76 & 3.39 & 97.57 & 0.0363 & 1 & 4.496 & 4.541 & 5.718 & 5.719&	4.534	&	4.574\\
0.2 & 3193 & -0.7 & 3.4 & 102.6 & 0.0041 & 1 & 4.361 & 4.389 & 5.56 & 5.561     &	4.405	&	4.424\\
0.25 & 3242 & -0.61 & 3.43 & 110.47 & 0 & 1 & 4.16 & 4.163 & 5.326 & 5.327      &	4.201	&	4.198\\
0.3 & 3281 & -0.53 & 3.46 & 118.1 & 0 & 1 & 3.982 & 3.966 & 5.125 & 5.126       &	4.016	&	4.001\\
0.35 & 3318 & -0.46 & 3.47 & 125.44 & 0 & 1 & 3.82 & 3.787 & 4.944 & 4.945      &	3.844	&	3.823\\
0.4 & 3357 & -0.39 & 3.48 & 132.28 & 0 & 1 & 3.672 & 3.623 & 4.779 & 4.78       &	3.696	&	3.662\\
0.45 & 3388 & -0.34 & 3.49 & 138.4 & 0 & 1 & 3.548  & 3.486 & 4.64 & 4.641      &	3.575	&	3.528\\
0.5 & 3427 & -0.28 & 3.5 & 144.06 & 0 & 1 & 3.437 & 3.353 & 4.508 & 4.509       &	3.463	&	3.402\\
0.57 & 3485 & -0.21 & 3.52 & 151.93 & 0 & 1 & 3.293 & 3.174 & 4.33 & 4.331      &	3.311	&	3.231\\
0.6 & 3507 & -0.17 & 3.52 & 155.92 & 0 & 1 & 3.224 & 3.094 & 4.25 & 4.251       &	3.242	&	3.154\\
0.62 & 3525 & -0.15 & 3.52 & 157.89 & 0 & 1 & 3.182 & 3.048 & 4.2 & 4.201       &	3.198	&	3.106\\
0.7 & 3593 & -0.08 & 3.53 & 164.8 & 0 & 1 & 3.036 & 2.887 & 4.026 & 4.027       &	3.042	&	2.936\\
0.75 & 3632 & -0.04 & 3.54 & 169.47 & 0 & 1 & 2.945 & 2.789 & 3.919 & 3.92      &	2.939	&	2.831\\
0.8 & 3670 & 0 & 3.55 & 173.8 & 0 & 1 & 2.862 & 2.701 & 3.821 & 3.822           &	2.844	&	2.733\\
0.85 & 3705 & 0.04 & 3.55 & 178.34 & 0 & 1 & 2.78 & 2.613 & 3.726 & 3.727       &	2.748	&	2.638\\
0.9 & 3752 & 0.03 & 3.61 & 171.83 & 0 & 1 & 2.823 & 2.654 & 3.748 & 3.749       &	2.771	&	2.660\\
0.95 & 3776 & 0.11 & 3.56 & 186.04 & 0 & 1 & 2.633 & 2.461 & 3.555 & 3.556      &	2.574	&	2.466\\
1 & 3810 & 0.13 & 3.58 & 187.82 & 0 & 1 & 2.586 & 2.412 & 3.495 & 3.496         &	2.514	&	2.406\\
1.05 & 3841 & 0.17 & 3.57 & 193.6 & 0 & 1 & 2.497 & 2.322 & 3.396 & 3.397       &	2.415	&	2.307\\
1.1 & 3871 & 0.21 & 3.57 & 197.99 & 0 & 1 & 2.425 & 2.25 & 3.314 & 3.315        &	2.331	&	2.225\\
1.15 & 3899 & 0.23 & 3.58 & 201.14 & 0 & 1 & 2.369 & 2.193 & 3.248 & 3.248      &	2.262	&	2.158\\
1.2 & 3978 & 0.28 & 3.58 & 203.89 & 0 & 1 & 2.28 & 2.104 & 3.136 & 3.136        &	2.145	&	2.044\\
1.3 & 4098 & 0.35 & 3.6 & 208.83 & 0 & 1 & 2.141 & 1.967 & 2.962 & 2.963        &	1.952	&	1.865\\
1.4 & 4197 & 0.42 & 3.6 & 214.98 & 0 & 1 & 2.006 & 1.836 & 2.797 & 2.797        &	1.754	&	1.691\\
\enddata


\tablecomments{Version Oct.2016}


\end{deluxetable}

\begin{deluxetable}{cc}
\tablewidth{0pt}
\tablecaption{Spectral Type vs. Effective Temperature relation according to \cite{2016ApJS..225...10F}
\label{tab:Faherty}
}
\tablehead{
\colhead{Spectral Type} & 
\colhead{$T_{eff}$(K)} 
}
\startdata
L0    &2173\\
L1 & 1983   \\
L2 & 1809   \\
L3  & 1649  \\
L4  & 1504     \\
\enddata
\end{deluxetable}

\section{Empirical Ridge Line}
In this appendix we describe the process leading to our empirical isochrone, i.e., a mean locus in our near infrared CMD where all our observations would, on average, lie after extinction correction.
We refer to this locus as ridge line. Several factors have to be taken into account to build a ridge line that can be compared to a theoretical isochrone, namely membership, differential extinction, distance, binaries and photometric errors.

\begin{figure}[t]
\begin{center}
\includegraphics[width=0.49\textwidth,trim={0 0.5cm 0 1.5cm}]{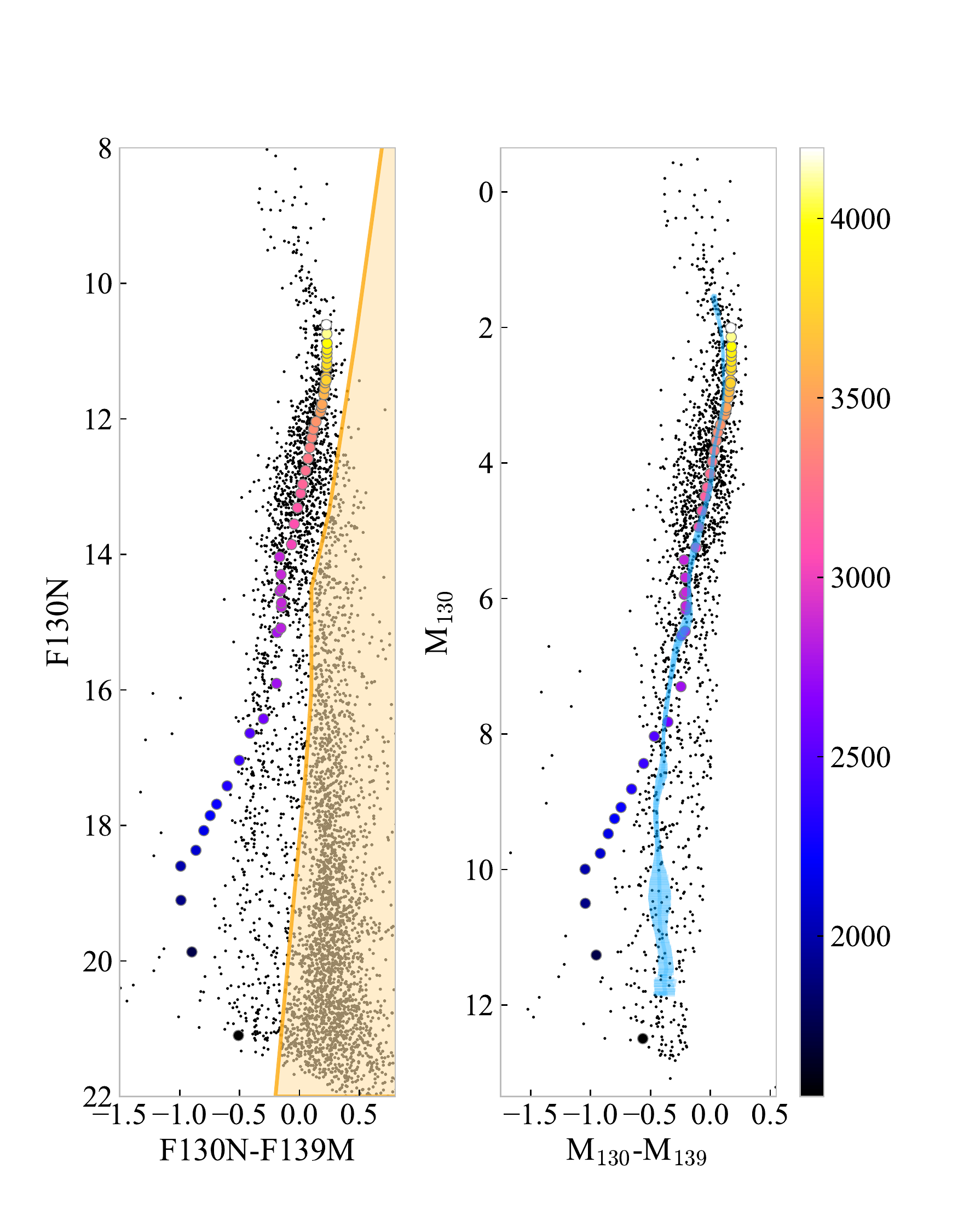}
\caption{Left: our near-infrared CMD, with superimposed a 1~Myr isochrone at a distance of 403~pc reddened assuming the median visual extinction of the cluster members ($A_V=2.18$) as derived from \cite{2011AA...533A..38S} map, and assuming A$_{130}$/A$_{V}$ = 0.264 and A$_{139}$/A$_{V}$ = 0.241. 
The orange area defines the exclusion region for background objects. Right:
 absolute magnitudes ($d = 10$~pc) and colors of the stars considered as members, and the 1~Myr isochrone without reddening. The blue stripe represents the empirical ridge line, with variable thickness representing its 1-sigma uncertainty.  The dots describing the isochrone are color-coded according to the linear effective temperature scale shown at the right side of the figure.
\label{fig:ridge_line}}
\end{center}
\end{figure}

\begin{deluxetable}{cc}
\tabletypesize{\normalsize}
\tablecaption{Edges of the segmented line describing the exclusion limit for background sources. Sources redder than the segmented line in the table are considered background for the purpose of defining the empirical ridge line\label{tab:poly}.}
\tablehead{
\colhead{$m_{130}-m_{139}$} & 
\colhead{$m_{130}$} 
}
\startdata
1    & 4 \\
0.47 & 10.80   \\
0.25 & 13.35   \\
0.1  & 14.5       \\
0.1  & 16.0     \\
0.05 & 17.25   \\
-0.2 & 22 \\
\enddata
\end{deluxetable}

\paragraph{Membership}
Using our CMD it is easy to separate the majority of low mass members of the ONC from background stars. We trace a line in the CMD that approximately follows the distribution of ONC sources, shifted to bluer colors to separate the large population of objects that clearly stand out as non-members (Fig.~\ref{fig:ridge_line}, left panel). We shall only consider in our subsequent analysis the objects that fall to the left of the line. This leaves us with 1,830 sources. The edges of the adopted segmented line are given in Table~\ref{tab:poly}.

\paragraph{Accounting for Differential Extinction}
Given our assumption that the discrepancy between models and data is intrinsic to the atmosphere treatment, it is preferable to derive the correction in the CMD of absolute magnitudes. In order to do this we have to subtract the extinction in each band from the photometry of our sources.
\cite{2011AA...533A..38S} derived a map of the extinction in the region. We use their map to remove extinction with two caveats. First, we have to assume a reddening law to connect the $A_V$ from the map to the extinction in each band. We use the \cite{1989ApJ...345..245C} law, with $R_V =3.1$, and the relations for $A_{130}/A_{V}$ and $A_{139}/A_{V}$ presented in Eq. \ref{eq:Alambda1} and \ref{eq:Alambda2}.
Second, the A$_V$ values from \cite{2011AA...533A..38S} are spatial averages over an area of few square arcmins. To assume for all nearby sources the same average extinction would ignore the scatter between individual sources present in the original data. Therefore, we draw instead an A$_V$ value for each source from a log-normal distribution. The $(\mu,\sigma)$ parameters of the distribution are set to: 
\begin{eqnarray}
\mu &=& \log(\mathrm{A}_{V,\mathrm{map}}) -0.5 \log(1+\alpha^2)\\
\sigma &=& \sqrt{\log(1+\alpha^2)},
\end{eqnarray}
where $A_{V,\mathrm{map}}$ is the extinction value from the \cite{2011AA...533A..38S} map at the position of that source.
The log-normal distribution arises naturally when describing the column density distribution of the turbulent ISM \citep{2007ApJ...665..416K}, and we thus use it to parameterize the distribution of extinction.
The parameter $\alpha$ is equal to the ratio between the standard deviation and the mean of the distribution, thus an $\alpha$ of 0.1 means that the distribution's width is about 10\% of the distribution mean.
We adopt $\alpha = 0.3$, which in turn gives $\sigma =0.29$, similar to what expected from ISM simulations \citep[see][]{2017NJPh...19f5003K}.

\paragraph{Distance}
As described above for the extinction, distance needs to be accounted for in order to compare data and models in absolute magnitudes. For the ONC we adopt a distance of 403~pc, determined by \cite{2019ApJ...870...32K} on the basis of Gaia DR2 data and APOGEE-2 near-IR spectroscopy. 
A recent estimate by \citep{2017ApJ...834..142K}, based on Gaia DR2 data and APOGEE-2 near-IR spectroscopy, provides as slightly shorter distance, $386\pm3$~pc. For comparison, \cite{2004ApJ...610.1045S} assumed 480~pc while \cite{2012ApJ...748...14D} adopted $d = 414\pm7$ pc from \cite{2007A&A...474..515M}. The recent studies utilizing GAIA data seem to confirm a distance close to 400~pc, are their discrepancy translates to a { systematic} uncertainty of $\simeq$ 0.05 mag, irrelevant for our discussion.

\paragraph{Binaries}
Given a set of points in a CMD, the method for computing the ridge line  is described in \cite{2016ApJ...823...18C}. These authors describe how such a method is able to trace the mode of the population, and it is not affected by the presence of the ``parallel sequence'' of binaries that could bias the ridge-line construction.

\paragraph{Photometric errors}
We account for the effect of photometric errors on the description of the ridge line by bootstrapping a set of catalogs, each of which contains the same number of stars as the original catalog members (i.e., the objects outside the orange region of Fig.~\ref{fig:ridge_line}, left). 

\paragraph{Single Iteration}
We create a catalog  by first creating a bootstrapped list of 1,830 sources, the number of non-background sources that may be preliminarily associated to the cluster population.
This means that we extract 1,830 sources among the 1,830 existing ones, allowing for repetitions. We then 
subtract from each observed data point the distance modulus and extract a random value for its $A_V$ from the individual log-normal distributions described above in this Section and subtract the extinction in each band.
We then obtain a single-iteration ridge line from this catalog using the method described in details in \cite{2016ApJ...823...18C}.

The method consists of estimating the source density in the CMD using a Kernel Density Estimation. We then find the maximum of such a density distribution (the peak of this ragged density landscape),
and move along the direction of minimum gradient, thus tracing the 
ridge.

\paragraph{Final Step}
We repeat the process for 5,000 bootstrapped catalogs.
At each iteration, different sources may be present in the catalog, and every time a source is present, it has a different random value of the extinction. At each iteration, the ridge line will be slightly different, thus accounting for all the uncertainties above.

We combine the results of the 5,000 iterations by binning the points from the ensemble of ridge lines in $m_{130}$ absolute magnitude bins. We derive the average color of the combined ridge line as function of magnitude. We also derive the ridge line color error, i.e., the standard deviation of the color distribution within each bin. The result is shown 
in Fig.\ref{fig:ridge_line}, right panel.

\subsubsection{Empirical isochrone\label{Sec:EmpiricalIsochrone}}
\begin{figure}[t]
\begin{center}
\includegraphics[width=0.49\textwidth, trim={0.5cm 1cm 0.5cm 2.5cm},clip]{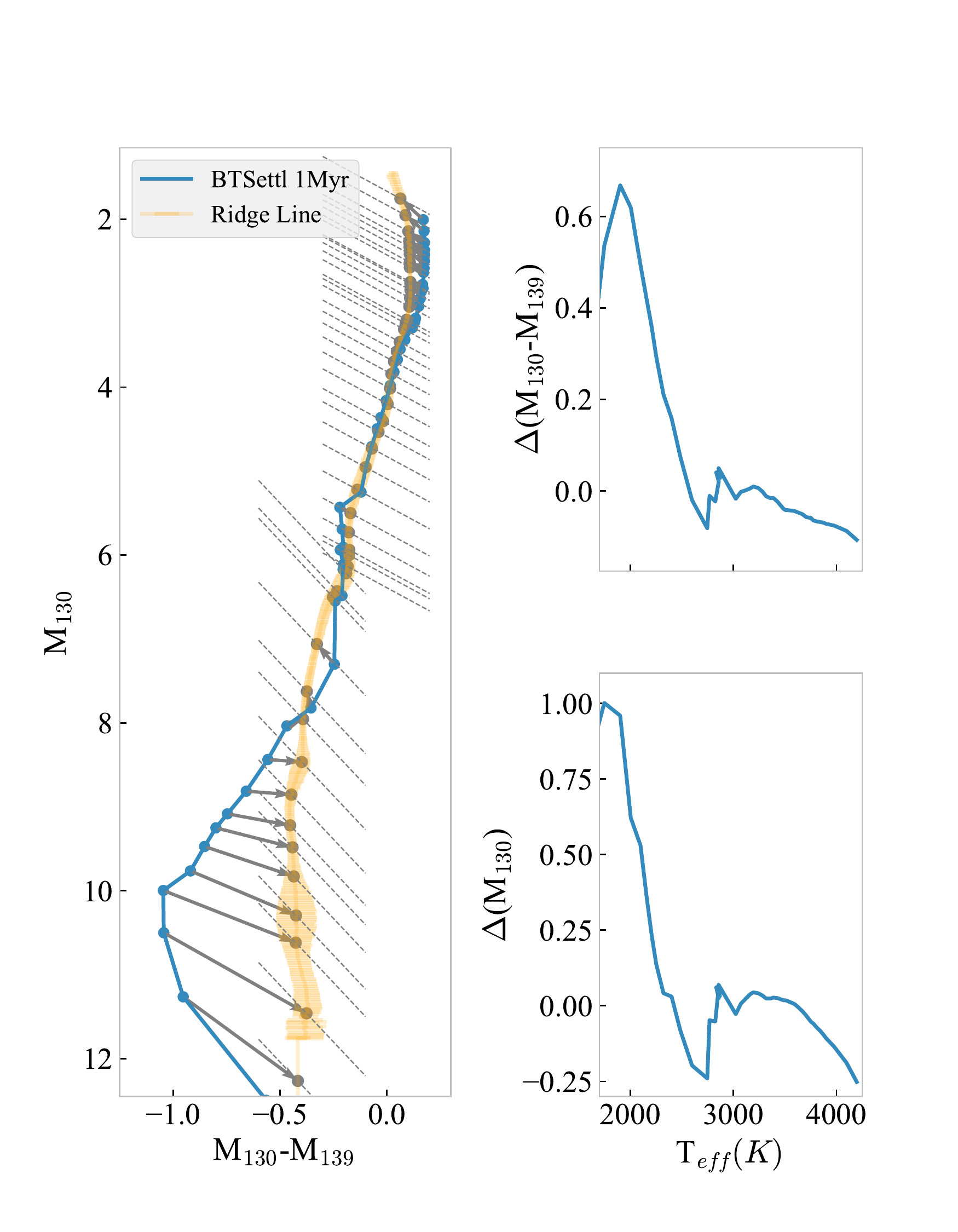}
\caption{Left: Empirical correction for each point along the isochrone. The dashed lines correspond to the constraints from the color-color diagram, Eq.~\ref{eq:2CD}.
Each point on the isochrone is shifted, to meet the grey points which represent the intercept between its constraint line (based on the above equations, for the F850LP value corresponding to that point) and the ridge line. Right: the resulting shifts in color (top panel) and magnitude (bottom panel) as a function of the temperature along the isochrone.\label{fig:correction}}
\end{center}
\end{figure}

Having determined the ridge line, we can derive the correction needed to adjust the theoretical isochrone to the data. 
For each mass value along the 1~Myr BT-Settl isochrone we use the appropriate version of Eq.~\ref{eq:2CD},
according to whether the mass is smaller or larger than $0.07~\Msun$.
We then use the corresponding value of $m_{850}$ for that mass, to derive a straight line in the $m_{130}$ vs ($m_{130}-m_{139}$) CMD. 
We find the intercept between this straight line and the ridge line; the intercept satisfies both constraints set by the color-color diagram (i.e. the $A_V=0$ lower envelop), and by the CMD ridge-line.
In terms of absolute magnitudes, we determine the shift in $M_{130}$ and $M_{130} - M_{139}$ color needed to move the point on the theoretical isochrone, for the mass in question, toward the corresponding dual-constraint point in the CMD.
We repeat this for all masses along the isochrone.
Finally, we linearly interpolate the shifts in magnitude and color as function of the effective temperature along the 1Myr BT-Settl isochrone and derive empirical corrections, $(\Delta_{M_{130}},\teff)$ and $(\Delta_{M_{130}-M_{139}},\teff)$. The process is illustrated in Figure~\ref{fig:correction}.

\begin{figure}[htb!]
\begin{center}
\includegraphics[width=90 mm,angle=0]{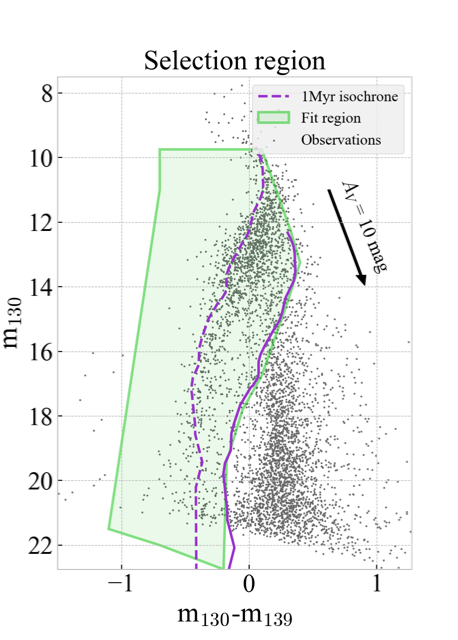}
\caption{Illustration of the area used to determine an extinction limited sample of low mass members of the Orion Nebula Cluster. \label{Fig:Selection_region}}
\end{center}
\end{figure}

\end{document}